\documentclass[twocolumn,trackchanges]{aastex701}

\begin{document}

\title{Energy Dependent Variability Properties of MAXI~J1820+070 as Observed by NICER and AstroSat}

\author{Mary Bosco}
\affiliation{Centre of Excellence in Astronomy and Astrophysics (CEAA), Department of Physics \& Electronics, Christ University, Bangalore 560029, India}
\email[show]{mary.bosco@res.christuniversity.in}  

\author{V. Jithesh} 
\affiliation{Centre of Excellence in Astronomy and Astrophysics (CEAA), Department of Physics \& Electronics, Christ University, Bangalore 560029, India}
\email{jithesh.v@christuniversity.in}

\author{Akash Garg}
\affiliation{Inter-University Centre for Astronomy and Astrophysics, Post Bag 4, Ganeshkhind, Pune 411007, India}
\affiliation{Persistent Systems Ltd., Pune, Maharashtra 411006, India}
\email{akashgarg_16@yahoo.co.in}

\author{Ranjeev Misra}
\affiliation{Inter-University Centre for Astronomy and Astrophysics, Post Bag 4, Ganeshkhind, Pune 411007, India}
\email{rmisra@iucaa.in}

\begin{abstract}

We present broadband spectral and temporal studies of the black hole X-ray binary MAXI~J1820+070 using nearly simultaneous observations from NICER and AstroSat. The combined NICER/XTI and AstroSat/LAXPC spectra in the 0.7 - 30~keV band were fitted with a model comprising disk emission and a Comptonizing component. The source was found to be in the hard state in the first epoch, the high-soft state during the second, and the quiescent state during the third. The power density spectrum of epoch 1 was modeled using multiple Lorentzians, confirming three prominent features at $\sim 50$~mHz, $\sim 84$~mHz, and $\sim 3$~Hz. The energy-dependent fractional rms and time lag spectra corresponding to these features were modeled, accounting for variations in disk and coronal parameters, as well as the time delays between them. The modeling reveals that perturbations at all three frequencies in the epoch~1 originate in the disk and propagate inward, with travel timescales that are strongly frequency-dependent. The variations in the coronal parameters are larger than those in the disk, indicating that the corona plays a dominant role in shaping the observed variability in epoch 1. For epoch~2, energy-dependent variability properties were studied at $\sim$0.1~Hz. The broad energy coverage afforded by the combination of NICER and AstroSat data proved essential in getting a more complete physical picture of the variability mechanism.

\end{abstract}



\hypersetup{pageanchor=true}

\section{Introduction} 

X-ray binaries (XRBs) are gravitationally bound systems of two stars, one being a compact object and the other a companion star \citep[main sequence, giant or super-giant;][]{Remillard_2006}. Accretion onto black hole X-ray binaries (BHXRBs) emits radiation across energies, with most BHXRBs showing spectral changes from hard to soft states via intermediate ones \citep{Homan_2005, Belloni_2011}. In the Hard State (HS), the spectrum is dominated by non-thermal emission from a hot, optically thin corona \citep{Shapiro_1976}, while the Soft State (SS) is marked by soft thermal emission from a geometrically thin, optically thick disk \citep{Shakura_1973}. Intermediate states show contributions from both components.  BHXRBs show rapid timing variability, quantified using the Power Density Spectrum (PDS), obtained from the Fourier transform of the light curve \citep{Belloni_2009}. The PDS typically shows broadband noise and, at times, narrow peaks known as quasi-periodic oscillations \citep[QPOs; see examples, ][]{McClintock2004, belloni_2010fastvariability, Belloni_2016}. 
\\

\subsection{Timing lag models}
Energy-dependent features, such as fractional rms (frms) and time lag, have been shown to correlate with the physical conditions of the corona and the accretion disk, suggesting that the observed variability arises from distinct radiative processes. To understand the radiative origin of this variability, several models \citep{Lyubarskii, Misra_2000, Kotov_2001, Arevalo_2006, Ingram_2011b, misra_soma_2013, Karpouzas_2021, Bellavita_2022} have been proposed to interpret energy-dependent frms and time lag. The single-zone stochastic propagation model proposed by \cite{Barimaqbool_2019} describes the energy-dependent variability properties using a simple geometry and minimal parameters, such as the inner disk radius, the truncation temperature, and the inter-region lag. The model has explained the timing features in BHXRBs such as Cygnus~X-1 \citep{Barimaqbool_2019}, Swift~J1658.2–4242 \citep{Jithesh_2019}, MAXI~J1820+070 \citep{Mudambi_2020} and MAXI~J1348–630 \citep{Jithesh_2021}. A more sophisticated model by \cite{Garg_2020} utilizes variations in the physical parameters of the system corresponding to the disk blackbody and thermal Comptonization emission, to calculate the variability in each energy range. This scheme was initially applied to reproduce the energy-dependent frms and time lag spectra associated with the type-C QPO observed in GRS~1915+105 and was later extended to a broader sample of AstroSat LAXPC observations of MAXI~J1535–571 \citep{Garg_2022}. \cite{Nazma_2023} also implemented the same model to study the 2016 and 2017 outbursts from H1743–322 using AstroSat observations, successfully characterized the energy-dependent timing signatures of prominent QPOs and their harmonics. \cite{Tanenia_2024} presented a spectro-temporal analysis of BHXRB GX 339-4 during its 2021 outburst using simultaneous AstroSat and NICER observations, characterizing the energy resolved fractional rms amplitude and time lag spectra for two broad variability features across the 0.5 - 30~keV energy range. Using the simultaneous AstroSat and NICER observations of GRS~1915+105, \cite{Dhaka_2024} studied spectro-timing properties of the source and modeled the energy-dependent variability properties of the broad feature at $\sim 2$~Hz using the model proposed by \cite{Garg_2020}. The modeling suggests that the observed variation is driven by the coronal heating rate, which induces changes in the disk temperature and inner radius. \cite{arbind_2025b} conducted a spectro-temporal analysis of  BHXRB MAXI~J1803-298 using AstroSat/LAXPC and NICER observations, to identify Type-C and Type-B QPOs in multiple spectral states. The authors modeled the energy-dependent fractional rms and time lags as functions of accretion parameters and revealed that the dominant site of variability reverses dynamically between the disk and the corona over the course of the outburst. In addition, this formalism has been applied to neutron star (NS) X-ray binaries \citep{Sree_2024, Sree_2024b, arbind_2025}, and the model successfully explained the energy-dependent variability properties of NSXRBs.
\\
\subsection{MAXI~J1820+070}
 MAXI~J1820+070 was first identified by the All-Sky Automated Survey for SuperNovae (ASASSN) as an optical transient \citep[ASASSN-18ey;][]{Tucker_2018}. On 11 March, 2018, at 12:50 UT, the MAXI/GSC nova alert system detected the source at RA = 18h 20m 26s, Dec = +07° 02' 13" \citep{Kawamuro_2018}. The distance to this system was calculated by utilizing Gaia2 data by \cite{Gandhi_2019}  and reported to be $3.46^{+2.18}_{-1.03}$ kpc. The black hole mass has been estimated as $M_{\mathrm{BH}} = 8.48^{+0.79}_{-0.72}\,M_{\odot}$, based on spectral line shifts and ellipsoidal modulation of the companion star’s optical emission during quiescence \citep{Torres_2020}. \cite{Kara_2019} utilized NICER observations conducted in 2018 to demonstrate that the observed reduction in X-ray reverberation lags is driven by a shrinking corona rather than a retreating accretion disk. Using NICER observations, \cite{Ma_2023} captured the transition from type-C to type-B QPOs in MAXI~J1820+070. In addition, through the spectral and timing analysis using the time-dependent Comptonization model vkompth, they found that the corona undergoes a dramatic geometric transformation at the QPO transition, from a horizontally extended structure overlying the accretion disk in the hard intermediate state (HIMS) to a vertically extended, jet-like corona in the soft intermediate state (SIMS), coinciding with discrete radio jet ejections. A complete timing study of the source using Swift and NICER observations \citep{Stiele_2020} showed that QPOs were not present after October 2018 and that the Power Density Spectrum (PDS) could be modeled using two broad features.
 \\

AstroSat observations of MAXI~J1820+070 have been studied in the past to understand spectral and timing properties \citep{Mudambi_2020, Chakraborty_2020, Geethu_Prabhakar, Banerjee_2024}. \cite{Mudambi_2020} utilized the 2018 AstroSat observation and identified the source in the hard spectral state, reporting a QPO at 47 mHz. In addition, they studied the time lag and rms in the LAXPC energy band at three different frequencies and modeled them using a single-zone stochastic propagation model \citep{Barimaqbool_2019}. Broadband X-ray spectral analysis of AstroSat and contemporaneous NuSTAR observations using the reflection model indicates a vertically stratified inhomogeneous corona for the source \citep{Chakraborty_2020}. In addition, they estimated the mass of the black hole to be $6.7 - 13.9 \,M_{\odot}$. Using Swift/XRT, NICER, NuSTAR, and AstroSat observations, \cite{Geethu_Prabhakar} performed a most comprehensive spectral and temporal study of the source. They identified the source in the low/hard state during the first outburst, whereas in the second outburst, the source was dominated by disk emission. In both outbursts, broadband spectral modeling reveals a strong reflection signature, suggesting the presence of an evolving corona. Multi-wavelength spectral analysis using AstroSat SXT, LAXPC, CZTI, UVIT, NICER, and LCO observations provided constraints on the inner and outer geometries of the accretion flow in the hard and soft spectral states \citep{Banerjee_2024}. In the hard state, the accretion disk is truncated and replaced by two different Comptonization components. 
\\
\cite{kawamura_2023} extended the propagating-fluctuations framework of \cite{kawamura_2022} to Insight-HXMT observations of 2018 outburst of MAXI~J1820+070 in the 2.6--48~keV band. They showed that the derived propagation speed is more consistent with a standard and normal evolution (SANE) flow than a magnetically arrested disk \citep[MAD;][]{Narayan_2012}. Independently, \cite{Omama_2023} applied a time-domain differential cross-correlation technique to NICER data from the same outburst. They identified a soft lag (of $\sim$30~ms) between the two energy bands (0.5–1 keV and 1–10 keV), which is due to the reverberation lag between soft excess and Comptonized emission, while the detected hard lag ($\sim$3~s) is between the disk and the Comptonizing corona. Further, they studied the contraction of the inner disk radius in the hard state through the evolution of both hard and soft lags.
\\
In this work, we present a broadband spectro-temporal analysis of MAXI~J1820+070 using nearly simultaneous NICER and AstroSat observations. Here, we apply, for the first time, the \citet{Garg_2020} formalism to MAXI~J1820+070, modelling the energy-dependent fractional rms amplitude and time-lag spectra of three distinct variability features to constrain the respective contributions of the disk and corona to the observed variability. The combination of NICER/XTI and AstroSat/LAXPC data provides continuous coverage over the 0.7--50~keV energy range, which proves essential in constraining the scattering fraction, a parameter that remains poorly determined from hard-band data alone. Section~\ref{obs} describes the observations and data reduction procedures. The broadband spectral and timing analyses are presented in Sections~\ref{spec} and~\ref{time}, respectively. The modelling of energy-dependent variability properties is described in Section~\ref{energy_dependent}, and the results 
are discussed in Section~\ref{conclu}.

\section{Observations and Data Reduction}
 \label{obs}

\begin{figure}
    \centering
    \includegraphics[width=1.05\linewidth]{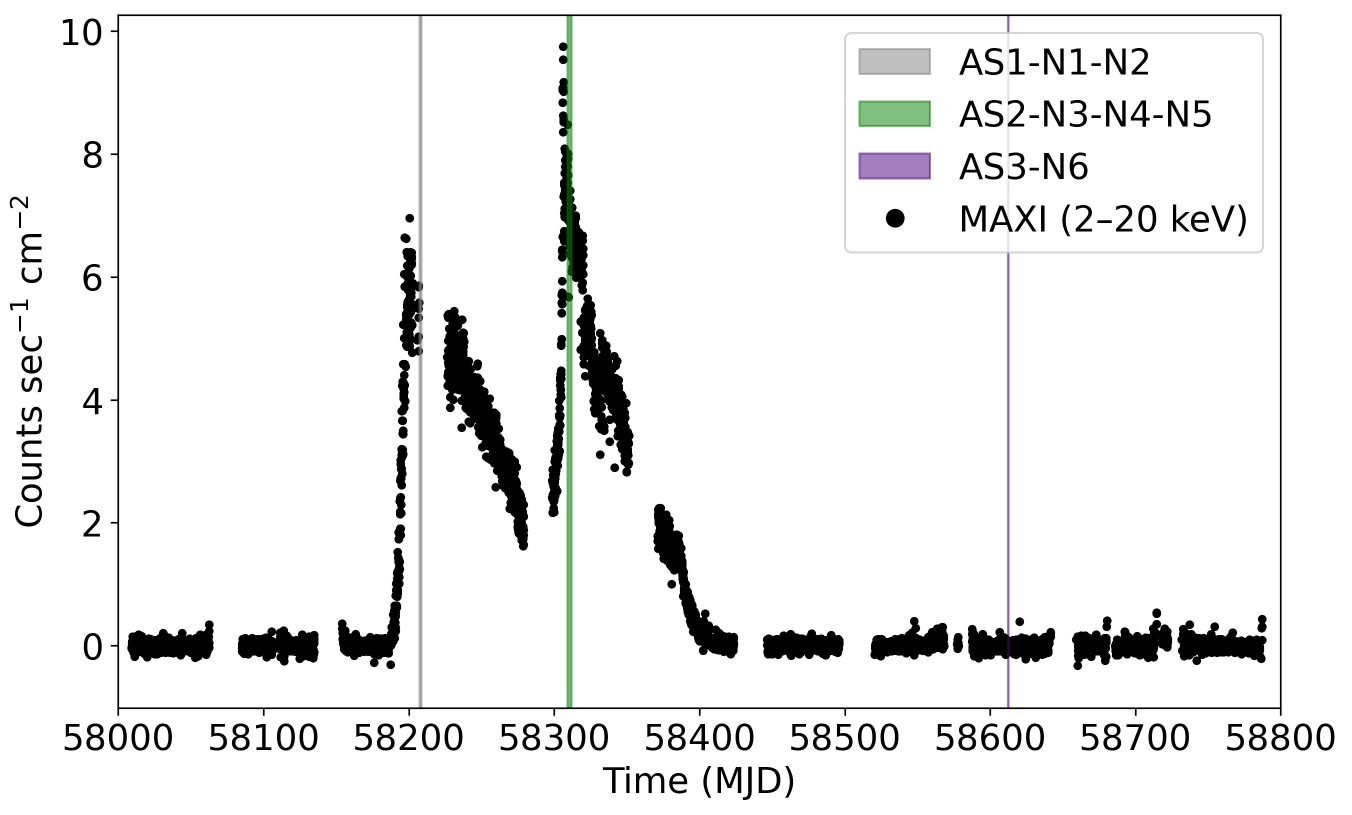}
    \caption{The one-day binned 2 - 20 keV MAXI light curve of MAXI~J1820+070 over the period of 2 years,
    with the vertical lines indicating the nearly simultaneous AstroSat and NICER observations that are being used.}
    \label{fig:MAXI_maxi18}
\end{figure}

AstroSat and NICER observed MAXI~J1820+070 on multiple occasions during the outburst. We selected nearly simultaneous AstroSat and NICER observations of the source for the broadband timing and spectral studies. The details of the observations are given in Table ~\ref{tab:data_used_j1820}. The epochs of these observations are indicated in the 2$-$20 keV MAXI light curve and are shown in Figure \ref{fig:MAXI_maxi18}. The first two AstroSat observations were conducted near the peak of the outbursts, and the third was in the quiescent state.

\begin{table*}
    \centering
	\caption{Observation log for MAXI~J1820+070}	\label{tab:data_used_j1820}
	\begin{tabular}{cccccccc} 
		
		\hline
		Epoch & Data & Obs ID & Date & Start MJD & End MJD & Exposure (ks)\\
		\hline
        Epoch 1&AS1 & T02\_038T01\_9000001994 & 2018 March 30–31 &58207.47 & 58208.59 & 97.34 \\
        &N1  & 1200120115 & 2018 March 30 & 58207.39 & 58207.48 & 7.68\\
        &N2  & 1200120116 & 2018 March 31 & 58208.35 & 58208.82 & 39.82 \\
        Epoch 2&AS2 & T02\_060T01\_9000002216 & 2018 July 10–12 & 58309.05 & 58311.78 & 237.20 \\
        &N3  & 1200120202 & 2018 July 10 & 58309.03 & 58309.99 & 83.48 \\
        &N4  & 1200120203 & 2018 July 10 & 58309.99 & 58310.63 & 54.37\\
        &N5  & 1200120204 & 2018 July 12 & 58311.01 & 58311.99 & 85.19 \\
        Epoch 3&AS3 & T03\_108T01\_9000002898 & 2019 May 09 & 58612.56 & 58612.66 & 32.03\\
        &N6  & 2200120331 & 2019 May 09 & 58612.20 & 58612.28 & 6.45\\
		\hline
	\end{tabular}
\end{table*}

\begin{figure}
\includegraphics[width=0.95\columnwidth]{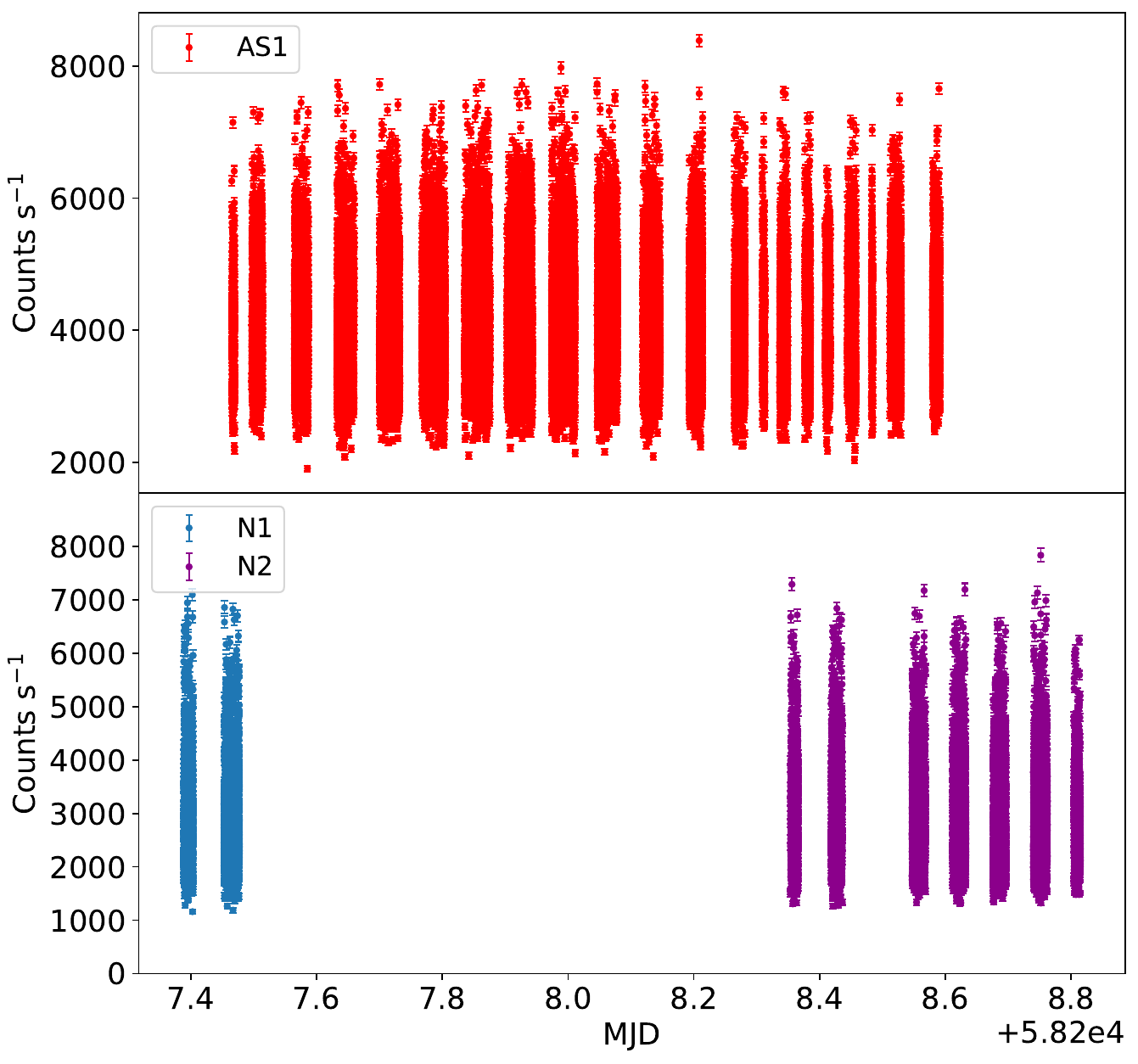}\\
\includegraphics[width=0.95\columnwidth]{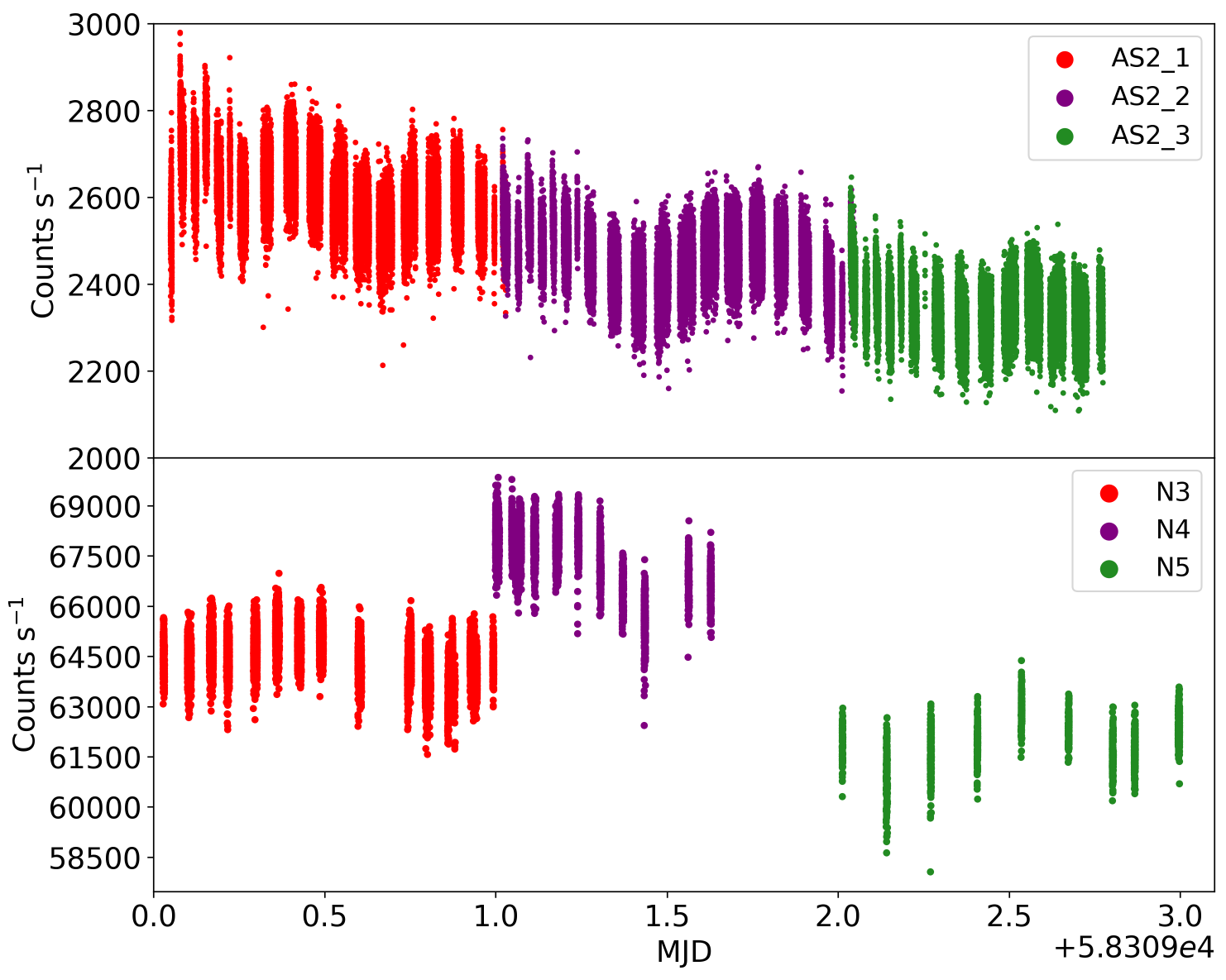}\\
\includegraphics[width=0.95\columnwidth]{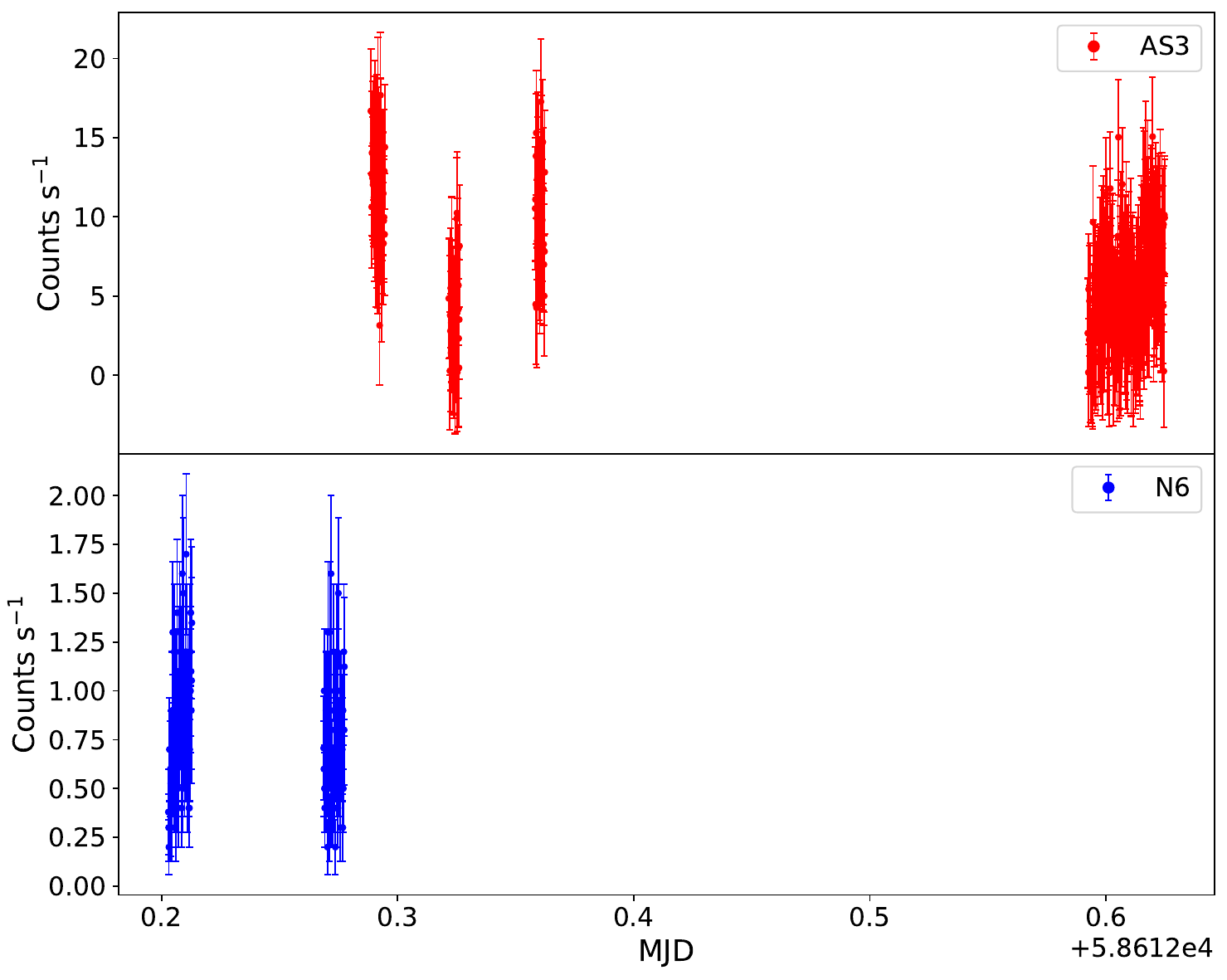}
\caption{Light curves extracted for the three different nearly simultaneous AstroSat/LAXPC and NICER/XTI observations. In each plot, the top panel represents the 3 - 50 keV background-subtracted light curve of MAXI~J1820+070 from AstroSat/LAXPC observation, and the bottom panel represents the 0.5 - 10 keV light curve from NICER/XTI observations. The light curves are binned at 10-second intervals.}
\label{lightcurves}
\end{figure}

\subsection{AstroSat}
\label{sec:AstroSat} 
We have used publicly available data of  MAXI~J1820+070 from \textit{AstroSat} \citep{Agrawal_2006, Singh_2014, Agrawal_2017, Singh_2022}. Among the various instruments onboard \textit{AstroSat}, we have utilized data from Large Area X-ray Proportional Counter (LAXPC).

\subsubsection{Large Area X-ray Proportional Counter}
LAXPC consists of three proportional counters (LAXPC10, LAXPC20 and LAXPC30) operating in the energy range of 3–80 keV with a temporal resolution of 10 µs \citep{Yadav_2016a,  Yadav_2016b, Agrawal_2017, Antia_2017}. However, among the three units, LAXPC10 and LAXPC30 exhibited abnormal behavior, including low gain and gas leakage. Moreover, LAXPC30 has been non-functional since March 8, 2018. Therefore, in this work, we used only LAXPC20 observations. We retrieved the Level 1 data from the ISSDC archive\footnote{\url{https://astrobrowse.issdc.gov.in/astro_archive/archive/Home.jsp}} and processed it to generate Level 2 event files using the latest version (as of 2022 August 15) of the LAXPC analysis software (LAXPCSOFT\footnote{\url{http://astrosat-ssc.iucaa.in/laxpcData}}). The observations were conducted in Event Analysis mode, which records the arrival time and energy of each detected photon. The Good Time Interval (GTI) files were generated using the LAXPC routine \texttt{laxpc\_make\_stdgti} \footnote{\url{http://astrosat-ssc.iucaa.in/uploads/threadsPageNew_SXT.html}}. Subsequently, the light curves and time-averaged energy spectra for all observations were generated using the \texttt{laxpc\_make\_lightcurve} and \texttt{laxpc\_make\_spectra} routines, respectively.

\subsection{NICER}
NICER \citep{Gendreau_2012} is an International Space Station (ISS) payload. NICER’s X-ray Timing Instrument (XTI) operates in the 0.2–12 keV range with an energy resolution of 85 eV at 1 keV. It offers high sensitivity, with an effective area exceeding 2000 cm² at 1.5 keV, and an excellent timing resolution of $\sim 100$ ns. Each NICER observation was processed using {\texttt nicerl2} \footnote{\url{https://heasarc.gsfc.nasa.gov/docs/software/lheasoft/help/nicerl2.html}} for standard and screening (NICER version: 2024-08-18\_V013 and calibration version: xti20240206). The light curves were extracted using \texttt{nicerl3\_lc}\footnote{\url{https://heasarc.gsfc.nasa.gov/docs/software/lheasoft/help/nicerl3-lc.html}}. Multiple NICER observations overlapping with AstroSat data (see Table \ref{tab:data_used_j1820}) were merged using \texttt{ftmerge}\footnote{\url{https://heasarc.gsfc.nasa.gov/docs/software/lheasoft/help/ftmerge.html}}. Spectral products were extracted with \texttt{nicerl3\_spect}\footnote{\url{https://heasarc.gsfc.nasa.gov/docs/software/lheasoft/help/nicerl3-spect.html}}, which incorporates systematic errors and quality flags and generates grouped spectra along with ARF, RMF, and background files. The \texttt{3C50} model \citep{remillard_3c50} was used for background estimation.

\section{Analysis and Results}
\label{ana}
\subsection{Spectral Analysis}
\label{spec}

We performed a joint spectral fitting of NICER (0.7-7 keV) and AstroSat LAXPC (3-30 keV) spectra  over a broad energy range of 0.7–30 keV using \textsc{xspec} (version 12.14.1). The spectral analysis was restricted to 30 keV owing to the increasing systematic uncertainties in the LAXPC background model at higher energies, which can significantly bias spectral parameter estimation. A 3 per cent systematic error was added to the LAXPC spectrum \citep{Misra_2017}, while for NICER, the systematic error is automatically added when running the \texttt{nicerl3\_spect} task. For modeling the spectrum, we start with a simple combination of \textit{diskbb} \citep{Mitsuda_1984} and \textit{thcomp} \citep{Zdziarski_2020} to account for the thermal disk emission and Comptonized emission, respectively. To account for the absorption of photons by the interstellar medium, we included the \textit{tbabs} \citep{WILMS_2000} component in \textsc{xspec}. The \textit{diskbb} model describes the thermal emission from the accretion disk, which has two parameters: the inner disk temperature ($kT_{\text{in}}$) and the disk normalization ($N_{\text{disk}}$). To model the hard X-ray component, we utilized the \textit{thcomp}, which is characterized by four parameters, including the power law photon index ($\Gamma$), electron temperature ($kT_{\rm e}$), scattering fraction ($f_{\rm sc}$), and redshift (z) of the source. The negative value of $\Gamma$ used in \textit{thcomp}\footnote{\url{https://heasarc.gsfc.nasa.gov/docs/software/xspec/manual/node322.html}} allows us to calculate the Thomson optical depth. \textit{thcompph} is a modified version of \textit{thcomp} that allows us to calculate the heating rate ($\dot{H}$) instead of the electron temperature during the spectral fitting as mentioned in \cite{Garg_2022}. To find $\dot{H}$, all other parameters were fixed to the values obtained from the previous spectral fit, and the best-fit value of the heating rate was then determined. In addition, a Gaussian component was added to account for the 6.4 keV iron line. The residuals seen in the spectra (at energies $\sim$ 1.1, 1.8 and 2.7 keV, see Table \ref{tab:spectral_params}) were modeled using the \textit{edge} components. The errors in the spectral parameters are calculated at a 90\% confidence level.

\begin{figure}
  \centering
  \includegraphics[width=1.2\columnwidth]{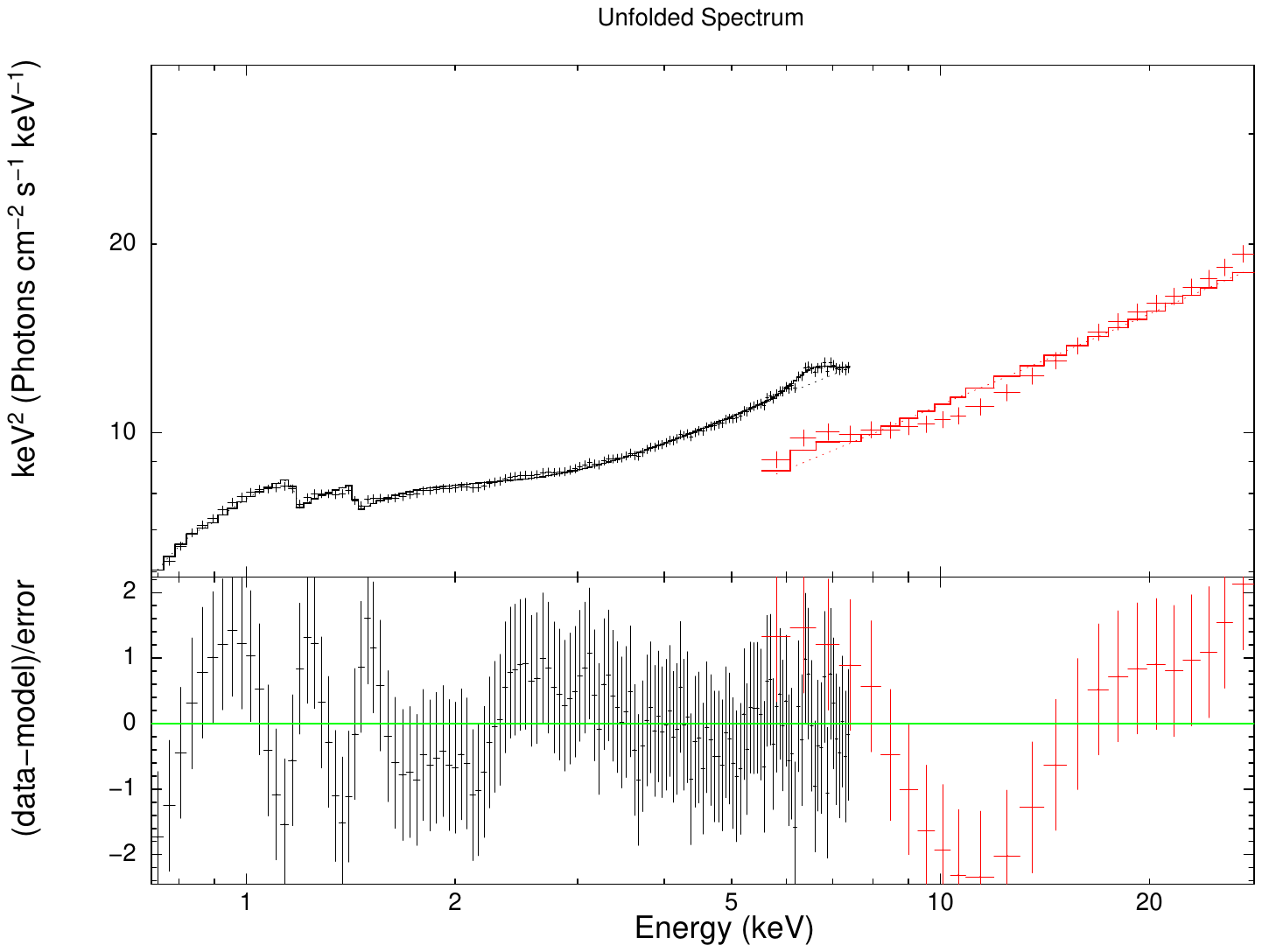}
   \caption{NICER (N1 and N2) and AstroSat LAXPC (AS1) spectra fitted with {\it constant*tbabs*edge*edge(thcomp*diskbb + gaussian)} model in the 0.7 - 30 keV energy range. The black and red data points represent NICER and LAXPC, respectively.}
\label{Spec_AS1}
\end{figure}

\begin{figure}

\includegraphics[width=1.2\columnwidth]{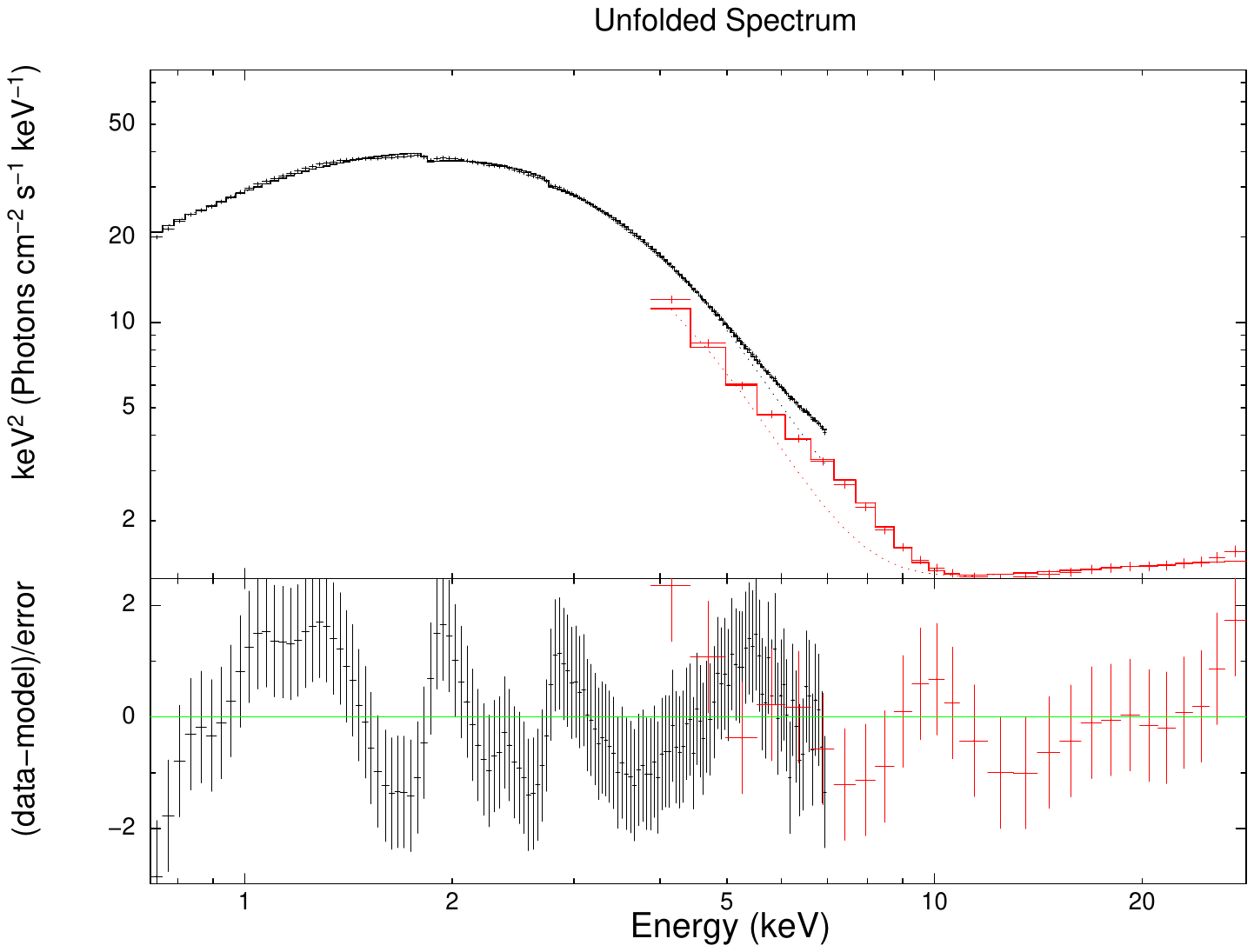}\\
\includegraphics[width=1.2\columnwidth]{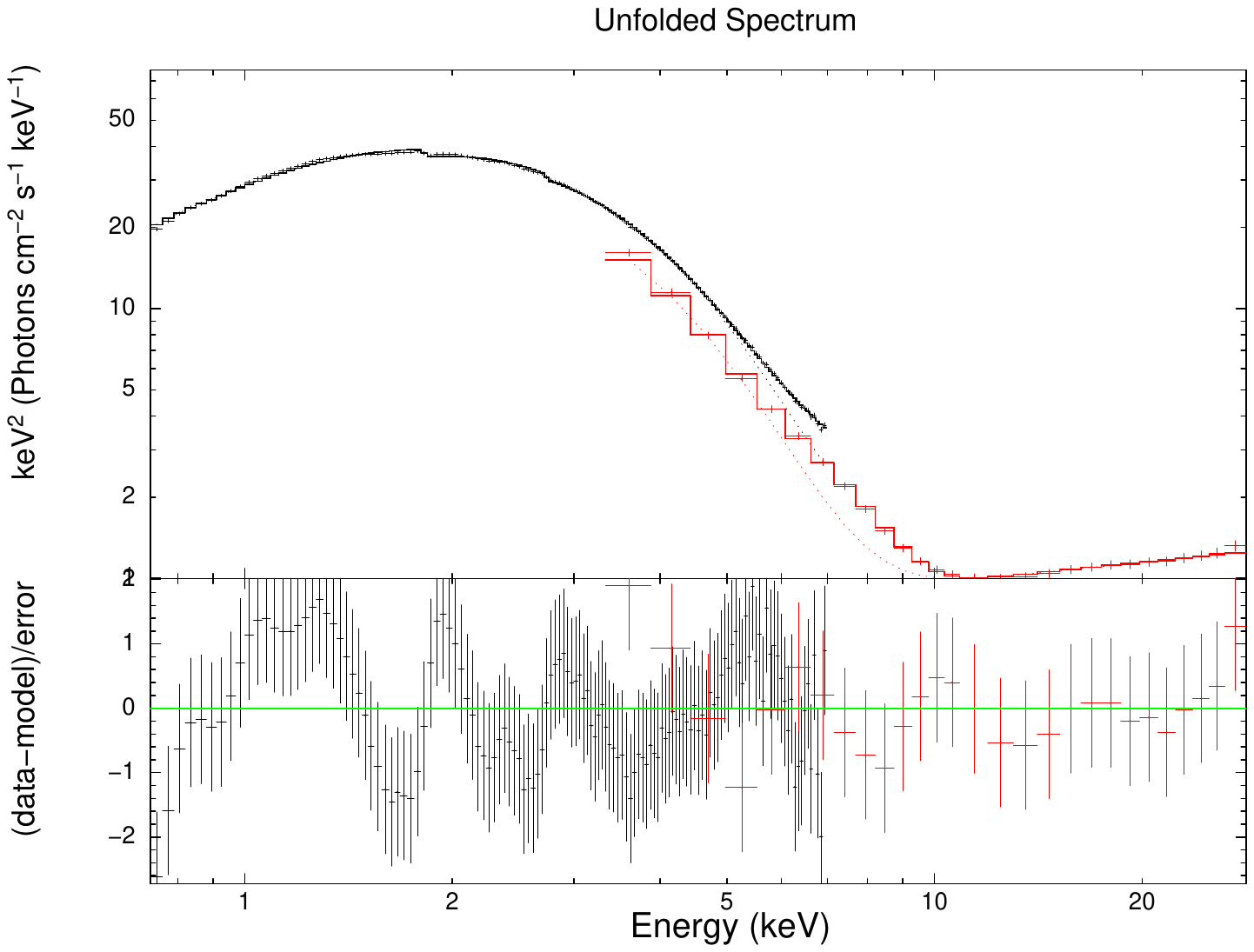}\\
\includegraphics[width=1.2\columnwidth]{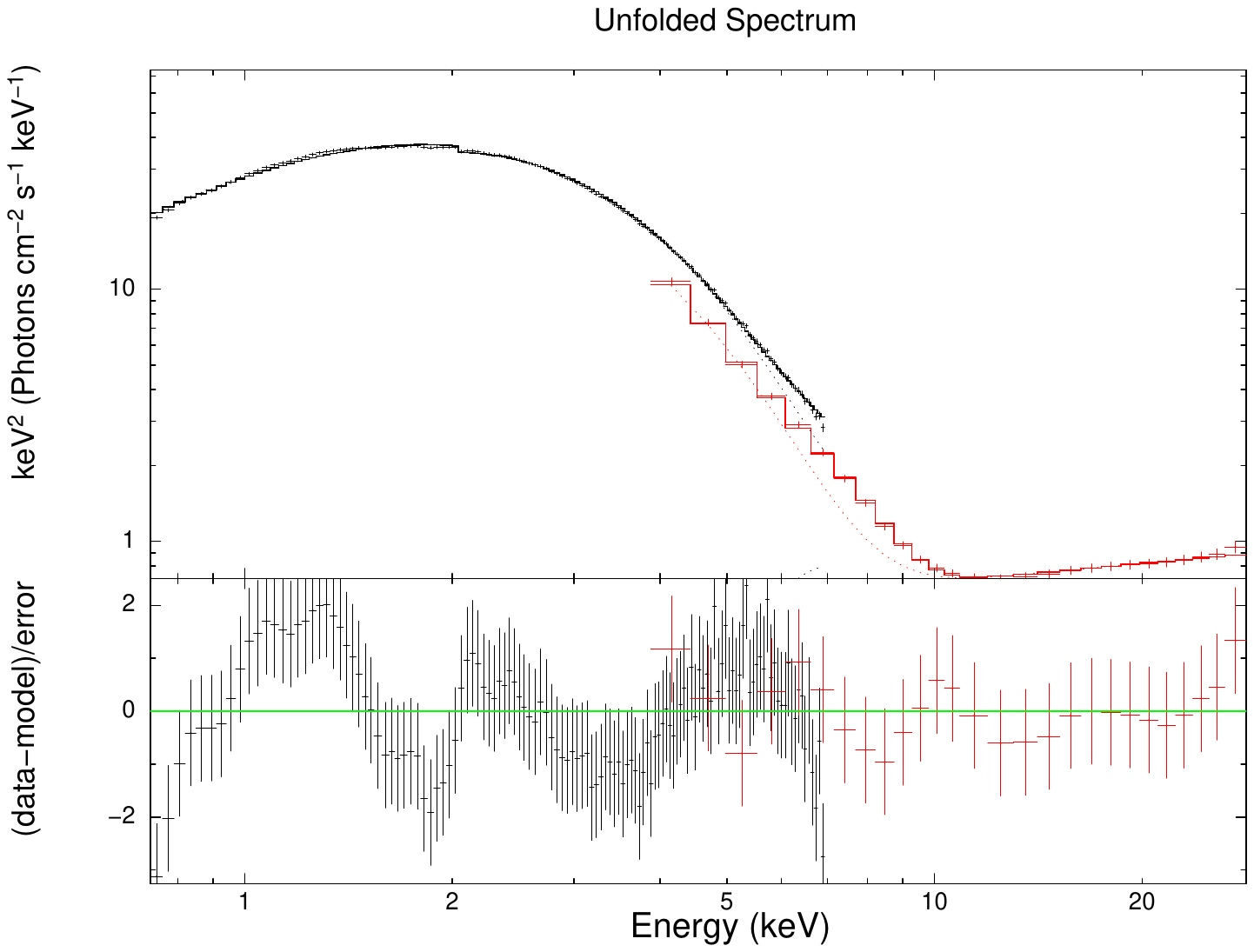}

\caption{Broadband energy spectrum of MAXI~J1820+070 obtained from the second epoch, which is split into three: AS2\_1+N3 (top), AS2\_2+N4 (middle) and AS2\_3+N5 (bottom) as discussed in section \ref{spec}. The spectra fitted with the model {\tt constant*TBabs*edge*edge(thcomp*diskbb + gaussian)}. NICER data (0.7 - 7 keV) and LAXPC data (3 - 30 keV) are denoted by black and red markers, respectively. The best-fit parameter values are summarized in Table~\ref{tab:spectral_params}.}
\label{spec_soft}
\end{figure}

 The first epoch observation of the source was carried out after the peak of the first outburst in 2018, which is reported as the failed outburst \citep{Geethu_Prabhakar}. In the first epoch, we have an AstroSat observation (AS1) and two NICER observations (N1 and N2; see the top panel of Figure \ref{lightcurves}). We modeled the NICER + AstroSat spectra using the model combination of {\it constant * tbabs * edge * edge (thcomp * diskbb + Gaussian)}. The coronal temperature is constrained with large uncertainty, the inner disk temperature is $\sim 0.38$ keV, the covering fraction is 0.54, and the optical depth is constrained to 2.86. The best-fit spectra are shown in Figure ~\ref{Spec_AS1} and the best-fit parameters are provided in Table \ref{tab:spectral_params}. We froze all parameters at their best-fit values and loaded the \textit{thcompph} model, which allows us to calculate the heating rate. The value of $\dot{H}$ was found to be 69.65 in the first epoch. The best-fit spectral parameters are used as input to the spectro-timing model \citep{Garg_2020}, which we used to fit the energy-dependent variability properties of the source (see Section \ref{energy_dependent}). The best-fit spectral parameters and spectral shape suggest that the source was in the hard state of BHXRBs. In the second epoch, the AstroSat observation (AS2) spans three days, and we have three NICER observations (N3, N4 and N5) in three consecutive days in 2018 July (see Table \ref{tab:data_used_j1820}). Thus, we split the AstroSat observation into three segments according to the NICER observations (see middle panel of Figure \ref{lightcurves}). We fitted the three broadband spectra from AstroSat and NICER using {\it constant * tbabs * edge * edge (thcomp * diskbb + gaussian)} and are shown in Figure \ref{spec_soft}. The best-fit parameters are given in Table \ref{tab:spectral_params}. In the spectral fitting, we could not constrain $kT_{e}$ and thus fixed it to 50 keV. The $n_{H}$ values marginally vary between (0.03 and 0.04) $ \times 10^{22}~cm^{-2}$, the covering fraction is $\sim 0.01$ - 0.03, the optical depth varies between 1.75 and 2.10, the inner disk temperature is $\sim 0.75$ keV and the disk normalization varies between {$3.52\times10^4$ and $5.09\times10^4$}. In the third spectrum (AS2\_3+N5) of the second epoch, we require only one edge component in the model. In the third epoch, the NICER observation was performed well before the AstroSat observation, and there was no overlap (see the bottom panel of Figure \ref{lightcurves}). In addition, the spectra from NICER and AstroSat were dominated by background. Hence, we have not included these observations in the further analysis.
 \\
We have noticed some residuals in the epoch 1 data. Thus, we modelled the spectrum using reflection model combination of (\texttt{constant*TBabs(edge*edge*relxilllpCp + diskbb)}. \cite{Geethu_Prabhakar} fitted the AstroSat (SXT and LAXPC) spectrum (AS1) with the same reflection model. The best-fit parameters obtained from our analysis are consistent with \cite{Geethu_Prabhakar}. Although the fit statistic of the reflection model is acceptable ($\chi^2/\mathrm{dof} =138.20/148$), the reflection modelling is not compatible with the energy-dependent variability framework of \citet{Garg_2020}. Thus, the baseline model, {\it constant * tbabs * edge * edge (thcomp * diskbb + Gaussian)}, is retained for the variability analysis.


\begin{table*}
\centering
\caption{Best-fit spectral parameters for different data groups fitted with
\textit{constant*tbabs*edge*edge(thcomp*diskbb + gaussian)}.}
\begin{tabular}{cllllll}
\hline
Component & Parameter & Unit & AS1+N1+N2 & AS2+N3 & AS2+N4 & AS2+N5 \\
\hline
constant   & factor                    & ---                 & $1.32^{+0.03}_{-0.03}$        & $1.44^{+0.04}_{-0.03}$        & $1.37^{+0.03}_{-0.03}$        & $1.41^{+0.03}_{-0.03}$        \\
TBabs      & $N_{\mathrm{H}}$          & $10^{22}$ cm$^{-2}$ & $0.05^{+0.01}_{-0.02}$        & $0.03^{+0.01}_{-0.01}$        & $0.04^{+0.01}_{-0.01}$        & $0.03^{+0.01}_{-0.01}$        \\
edge       & edgeE                     & keV                 & $1.17^{+0.03}_{-0.03}$        & $2.74^{+0.10}_{-0.09}$        & $1.81^{+0.03}_{-0.03}$        & $2.05^{+0.07}_{-0.12}$        \\
edge       & MaxTau                    & ---                 & $0.13^{+0.02}_{-0.02}$        & $0.06^{+0.01}_{-0.01}$        & $0.07^{+0.01}_{-0.01}$        & $0.06^{+0.01}_{-0.01}$        \\
edge       & edgeE                     & keV                 & $1.43^{+0.04}_{-0.04}$        & $1.82^{+0.10}_{-0.06}$        & $2.74^{+0.05}_{-0.11}$        & ---                            \\
edge       & MaxTau                    & ---                 & $0.11^{+0.02}_{-0.01}$        & $0.07^{+0.01}_{-0.01}$        & $0.05^{+0.01}_{-0.01}$        & ---                            \\
thcomp     & $\tau$                  & ---                 & $2.86^{+1.68}_{-1.33}$       & $1.90^{+0.01}_{-0.01}$       & $1.75^{+0.05}_{-0.05}$       & $2.10^{+0.14}_{-0.12}$       \\
thcomp     & $kT_{e}$                  & keV                 & $44.84^{+29.61}_{-19.11}$     & $50.00^{f}$                   & $50.00^{f}$                   & $50.00^{f}$                   \\
thcomp     & cov\_frac                 & ---                 & $0.54^{+0.14}_{-0.08}$        & $0.027^{+0.003}_{-0.003}$     & $0.017^{+0.002}_{-0.002}$     & $0.013^{+0.002}_{-0.002}$     \\
thcomp     & $z$                       & ---                 & $0.00^{f}$                    & $0.00^{f}$                    & $0.00^{f}$                    & $0.00^{f}$                    \\
diskbb     & $kT_{\mathrm{in}}$         & keV                 & $0.39^{+0.02}_{-0.01}$        & $0.75^{+0.01}_{-0.01}$        & $0.75^{+0.01}_{-0.01}$        & $0.75^{+0.01}_{-0.01}$        \\
diskbb     & norm                      & $10^{4}$            & $4.30^{+0.79}_{-0.78}$        & $1.28^{+0.04}_{-0.04}$    & $1.35^{+0.03}_{-0.03}$    & $1.28^{+0.03}_{-0.03}$    \\
gaussian   & LineE                     & keV                 & $6.51^{+0.07}_{-0.10}$        & $6.40^{f}$                    & $6.40^{f}$                    & $6.40^{f}$                    \\
gaussian   & Sigma                     & keV                 & $0.39^{+0.14}_{-0.11}$        & $1.28^{+0.09}_{-0.09}$        & $1.42^{+0.11}_{-0.12}$        & $1.40^{+0.11}_{-0.12}$        \\
gaussian   & norm                      & $10^{-2}$           & $1.19^{+0.004}_{-0.004}$      & $7.39^{+0.01}_{-0.01}$      & $5.47^{+0.01}_{-0.01}$      & $4.38^{+0.01}_{-0.01}$      \\
\hline
$\chi^2$/dof & & & $108.62/146$ & $136.09/151$ & $116.83/149$ & $155.24/145$\\
\hline
\multicolumn{7}{l}{$^{f}$ Parameter frozen during fitting.} \\
\end{tabular}
\label{tab:spectral_params}
\end{table*}

\subsection{Timing Analysis}
\label{time}

\begin{figure}
  \centering
  \includegraphics[width=1.2\columnwidth]{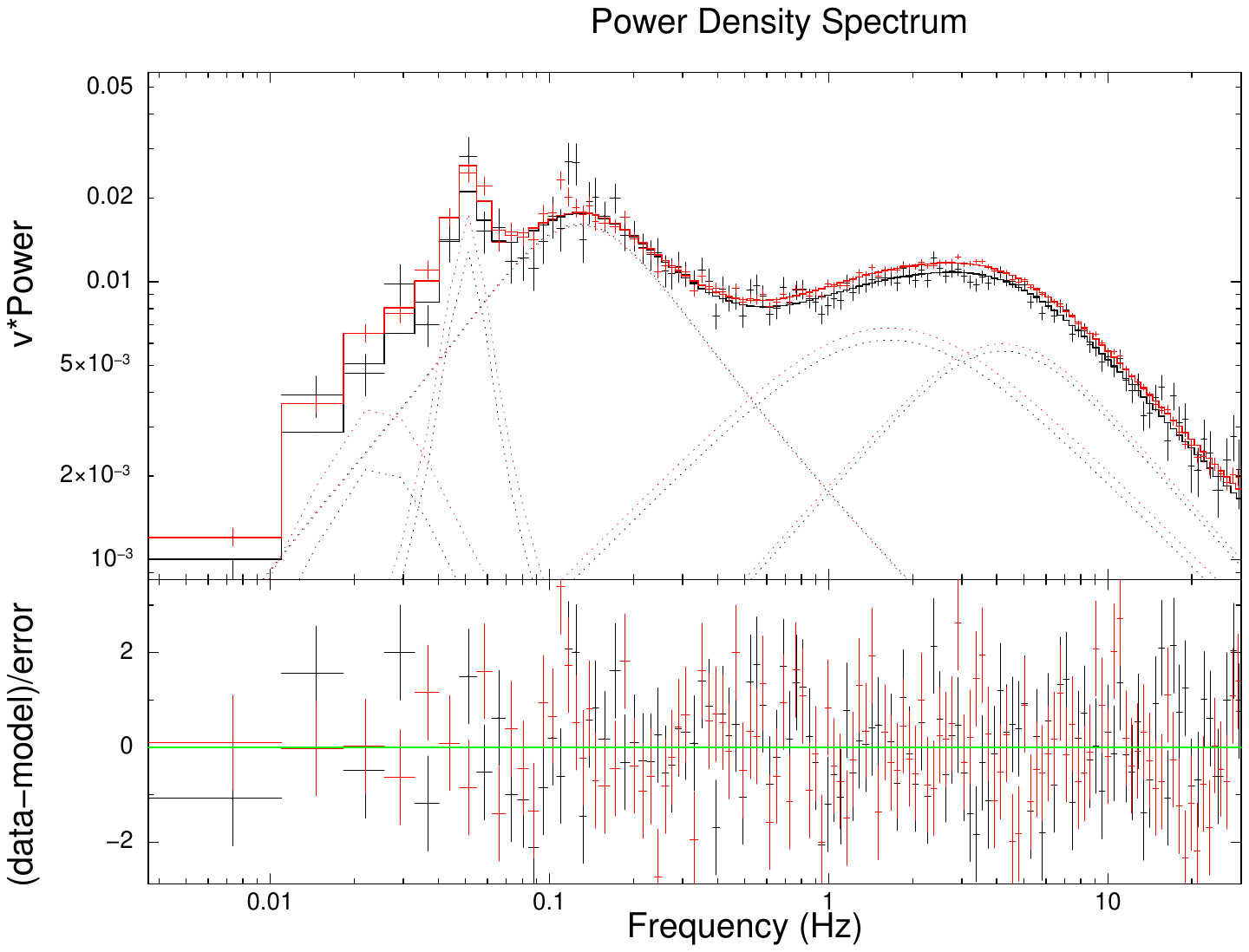}
   \caption{Fitted power density spectrum from AS1, N1 and N2 in the common energy range of 3-10 keV with five Lorentzians. Black and red data points represent LAXPC and NICER data, respectively.}
\label{PDS_AS1}
\end{figure}

The background-subtracted light curve of the source from NICER (0.5–10 keV) and AstroSat LAXPC (3–50 keV) extracted from all epochs is shown in Figure \ref{lightcurves}. A uniform time binning of 10 s has been adopted for all light curves. To quantitatively characterize the rapid X-ray variability components of the source, we derived the power density spectrum for LAXPC20 using the {\tt laxpc\_find\_freqlag} subroutine in the energy range of 3–10 keV (common energy range for NICER and AstroSat LAXPC). In the PDS, no significant variability was observed at higher frequencies; therefore, we used frequencies up to 10 Hz. In the first epoch, we observed two broad noise peaks and a sharp feature in the LAXPC PDS. We modeled the PDS with five Lorentzians. For NICER, we computed the PDS in the 3-10 keV energy range using the \texttt{nicer\_find\_freqlag} subroutine of the NICER\_RM\_software\footnote{Developed by Prof.~Ranjeev~Misra, Inter-University Centre for Astronomy and Astrophysics (IUCAA), Pune, which consists of subroutines similar to the LAXPC software to analyze NICER data.} The PDS features identified in the NICER observation match the LAXPC PDS. Thus, we fitted the NICER PDS with five Lorentzians, as in LAXPC, with the Lorentzian centroid frequency and width fixed to the values obtained from the LAXPC fit, while allowing only the normalization of each Lorentzian component to vary. The resulting fit is shown in Figure ~\ref{PDS_AS1}. The best-fit parameters derived from the combined fitting are listed in Table ~\ref{tab:pds_table}.  Three features were identified in the PDS, consistent with previous studies that reported the detection of a $\sim$ 47 mHz QPO \citep{Mudambi_2020}. The centroid frequencies of the two broad noise components are 84 mHz and 3 Hz, respectively, and the sharp feature is at 50 mHz.

\begin{figure}
    \centering
    \includegraphics[width=1.2\columnwidth]{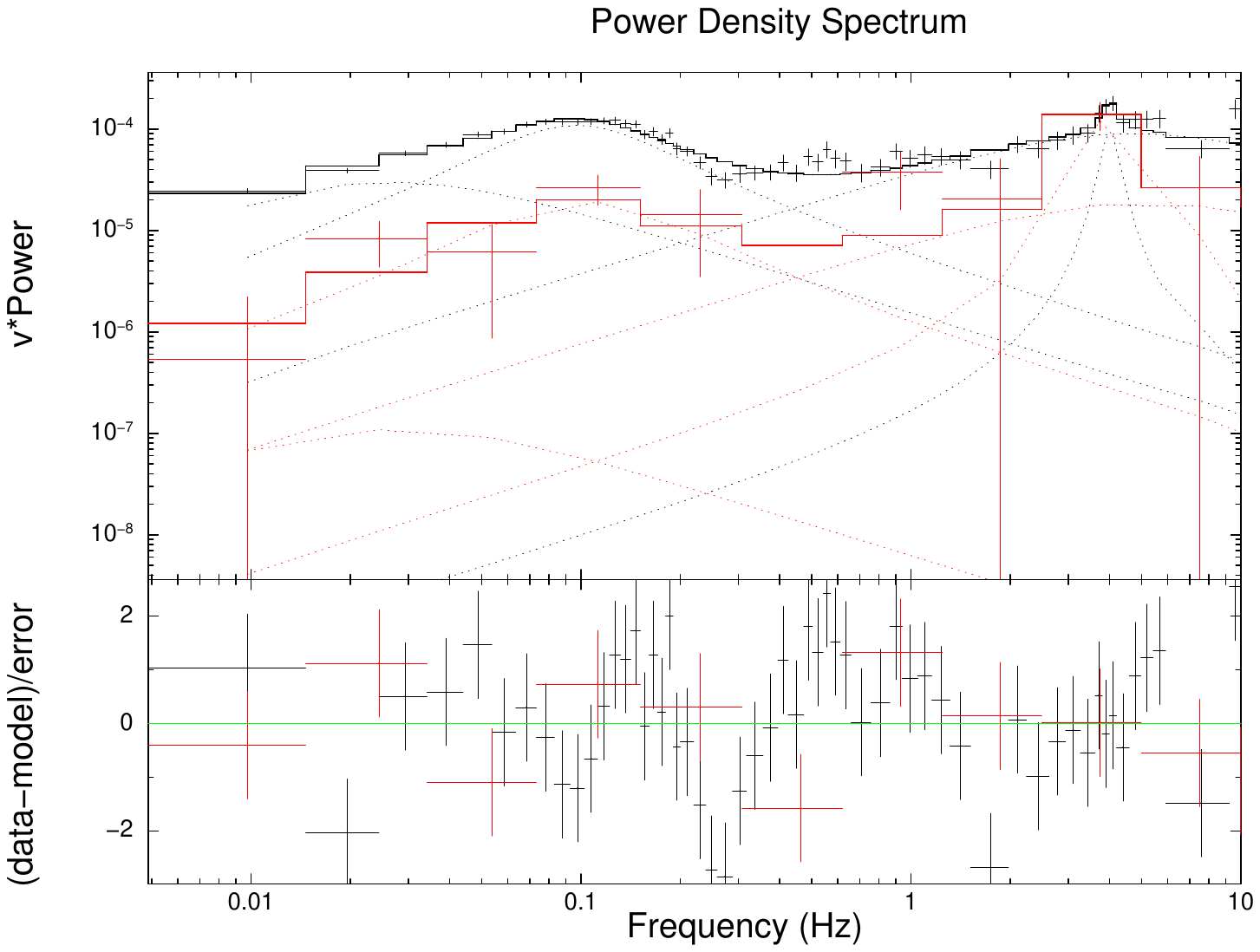}
    \caption{The PDS from the complete AS2 observation in the energy range of 3-10 keV energy band and N3, N4 and N5 extracted in the energy range of 0.5-4 keV modeled with 4 Lorentzians. Black and red data points represent LAXPC and NICER data, respectively. The best fit values are summarized in Table \ref{tab:pds_table}.}
    \label{PDS_AS2}
\end{figure}

In the second epoch, we initially extracted the PDS from three individual segments (in the 3-10 keV energy band) of NICER and LAXPC. However, individual power density spectra are weak; thus, we merged all three segments and extracted the PDS. Since the NICER power in 3-10 keV was very low, we have extracted the NICER PDS in 0.5-4 keV. The resulting LAXPC PDS (3-10 keV) is fitted with three Lorentzians and shown in Figure \ref{PDS_AS2}. In the PDS, we have detected a moderately broad feature at $\sim 3$ Hz. In the absence of a QPO, we extracted the fractional rms and time lags at different Nyquist frequencies (0.1, 1, and 10 Hz). Since the power was high around 0.1 Hz, we restricted our analysis to 0.1 Hz.

\begin{table*}
\centering
\caption{The best-fit parameters of PDS when modeled with multiple Lorentzian components.
The parameter uncertainties are at the 90 per cent confidence level. During fitting, the norm values are free for NICER data, while all other parameters are fixed to the best-fit values obtained from LAXPC data. Epoch 1 PDS required 5 Lorentzian components, while Epoch 2 required only 4.}
\begin{tabular}{cllll}
\hline
Component     & Parameter     & Unit & Value (Epoch 1) & Value (Epoch 2) \\
\hline

Lorentzian 1  & LineE         & Hz   &   $5.04^{+0.14}_{-0.13}\times10^{-2}$ &  $0.0^f$ \\
              & Width         & Hz   &  $1.39^{+0.41}_{-0.33}\times10^{-2}$  &  $9.57^{+1.99}_{-2.25}$  \\
              & norm (LAXPC)  & ---  &  $> 0.41 \times 10^{4}$ $^{\textbf{u}}$ &  $2.82^{+0.44}_{-0.47}\times10^{-4}$ \\
              & norm (NICER)  & ---  &  $7.86^{+0.18}_{-0.18}\times10^{-3}$   & $< 1.63\times 10^{-4}$ $^{\textbf{u}}$  \\

Lorentzian 2  & LineE         & Hz   &  $8.18^{+1.10}_{-1.23}\times10^{-2}$  &  $4.00^{+0.16}_{-0.14}$\\
              & Width         & Hz   &   $1.98^{+0.12}_{-0.07}\times10^{-1}$  &  $0.49^{+0.77}_{-0.20}$ \\
              & norm (LAXPC)  & ---  &   $3.43^{+0.31}_{-0.30}\times10^{-2}$ & $1.67^{+0.59}_{-0.86}\times10^{-4}$ \\
              & norm (NICER)  & ---  &  $3.39^{+0.29}_{-0.28}\times10^{-2}$   & $9.63^{+7.15}_{-7.02}\times10^{-5}$  \\

Lorentzian 3  & LineE         & Hz   & $0.00^{f}$                & $0.00^{f}$ \\
              & Width         & Hz   & $3.33^{+0.64}_{-0.68}$   & $5.01^{+4.47}_{-2.17}\times10^{-2}$ \\
              & norm (LAXPC)  & ---  &  $1.99^{+0.47}_{-0.45}\times10^{-2}$  & $9.43^{+7.88}_{-3.24}\times10^{-5}$ \\
              & norm (NICER)  & ---  &  $2.19^{+0.50}_{-0.48}\times10^{-2}$   &  $< 1.64 \times 10^{-5}$ $^{\textbf{u}}$\\

Lorentzian 4  & LineE         & Hz   &  $2.08^{+0.34}_{-0.24}\times10^{-2}$  & $7.28^{+1.94}_{-1.24}\times10^{-2}$  \\
              & Width         & Hz   &  $2.45^{+1.48}_{-0.80}\times10^{-2}$  & $0.13^{+0.01}_{-0.02}$  \\
              & norm (LAXPC)  & ---  &  $3.37^{+1.94}_{-1.62}\times10^{-3}$   & $1.98^{+0.39}_{-0.80}\times10^{-4}$ \\
              & norm (NICER)  & ---  &  $5.42^{+2.39}_{-1.51}\times10^{-3}$   & $3.88^{+1.98}_{-2.41}\times10^{-5}$  \\

Lorentzian 5  & LineE         & Hz   & $1.97^{+0.73}_{-0.49}$   & ---  \\
              & Width         & Hz   & $7.32^{+0.27}_{-0.39}$  & ---  \\
              & norm (LAXPC)  & ---  &  $1.33^{+0.49}_{-0.50}\times10^{-2}$  & ---  \\
              & norm (NICER)  & ---  &  $1.42^{+0.54}_{-0.54}\times10^{-2}$  & ---  \\

\hline
$\chi^2$/dof & & & $273.00/219$ & $95.59/53$ \\
\hline
\multicolumn{5}{l}{$^{f}$ Parameter frozen during fitting.} \\
\multicolumn{5}{l}{$^{u}$ Parameter is unconstrained; thus we report the upper/ lower limit.} \\
\end{tabular}
\label{tab:pds_table}
\end{table*}

\subsection{Energy-Dependent Variability Properties}
\label{energy_dependent}

We extracted energy-dependent variability properties, specifically the time lag and frms amplitude, over a broad energy range of 0.5-50 keV from NICER and AstroSat LAXPC observations. The 3 - 5 keV band was adopted as the reference for computing time lags for both the \textit{NICER} and \textit{LAXPC} data. To generate the time lag and frms, the LAXPC subroutine laxpc\_find\_freqlag  was used, which requires input parameters like frequency resolution
($\Delta f$) and the specific frequency (f) at which the energy-dependent properties were to be extracted. Using the command laxpc\_make\_lagrms\_pha, we create separate files containing phase lags and frms values, which are loaded into \textsc{xspec} for modeling. The observed energy-dependent variability was modeled using the framework proposed by \cite{Garg_2020}. In that model, the accretion flow comprises a thin, optically thick disk truncated at a radius beyond the ISCO, and a hot, optically thin corona of energetic electrons located close to the black hole. The parameters of the energy spectrum such as $kT_{\rm in}$, $N_{\rm disk}$, $\tau$, $kT_{\rm e}$, and $f_{\rm sc}$ were obtained from the combination of \texttt{diskbb} and \texttt{thcomp} model and input into the \cite{Garg_2020} model. To infer the physical properties of the system, we fixed these spectral parameters at their best-fit values. This modeling approach enables a direct mapping between spectral parameters and their physical counterparts. In particular, the inner disk temperature and the disk normalization correspond to the accretion rate and the inner disk radius, respectively. Similarly, $kT_{\rm e}$ determines the coronal heating rate ($\dot{H}$, from the thcompph fitting), while the scattering fraction and optical depth remain consistent between the spectral and physical representations. After obtaining the five input parameters, we assume that the observed variability is driven by small coordinated perturbations in these physical variables. The resulting energy-dependent timing observables are then derived from the complex linear response of the spectrum, $\Delta F(E)$. Specifically, the fractional rms is calculated as $\frac{1}{\sqrt{2}} \frac{|\Delta F(E)|}{F(E)}$, where $F(E)$ represents the steady-state spectral flux. The corresponding time lag is determined by $\text{argument of} \left( \Delta F(E_{\text{ref}})^{*} \Delta F(E) \right)$, representing the phase difference of the response relative to a reference energy band $E_{ref}$. 

\begin{figure*}
\centering

\includegraphics[width=0.32\textwidth]{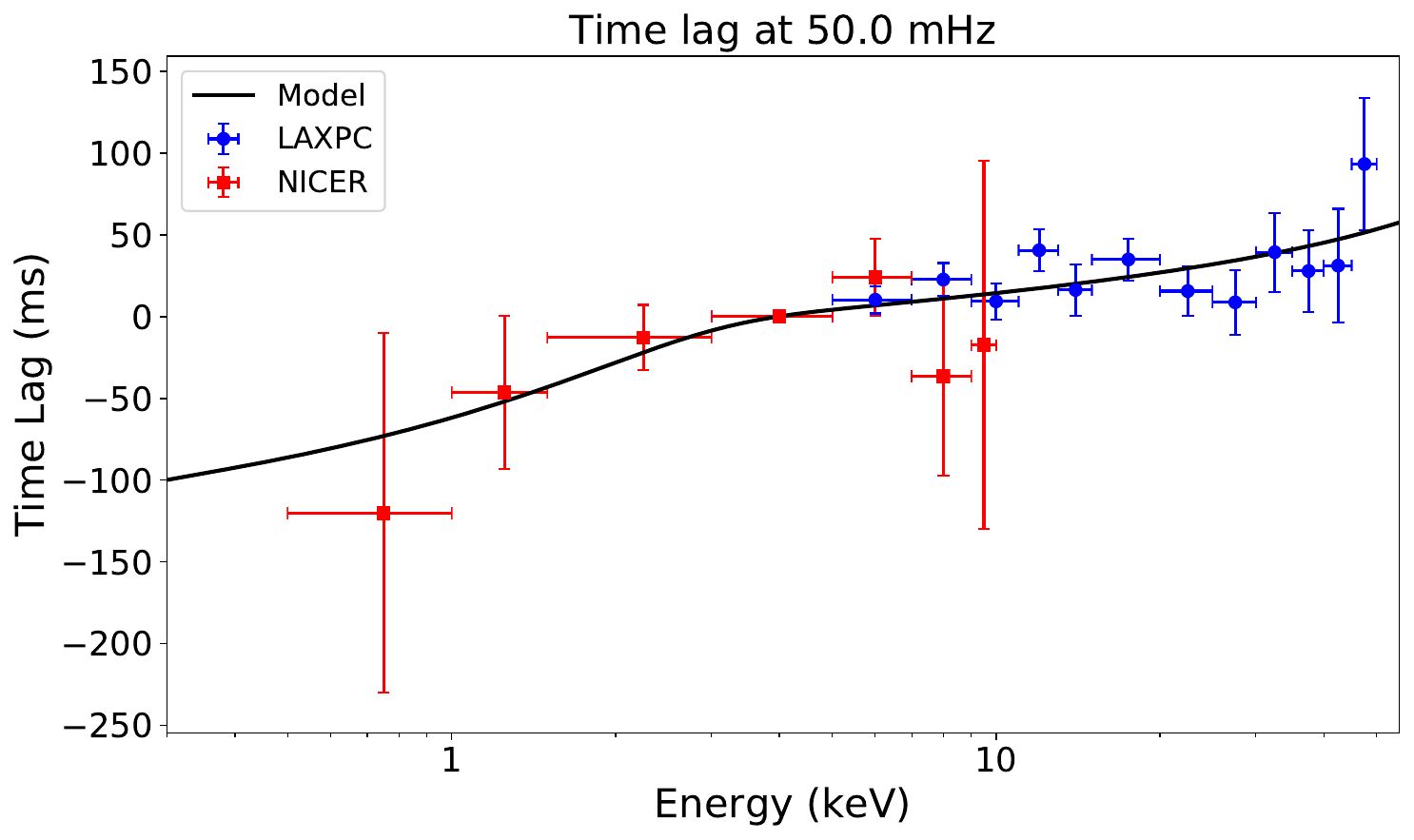}
\includegraphics[width=0.32\textwidth]{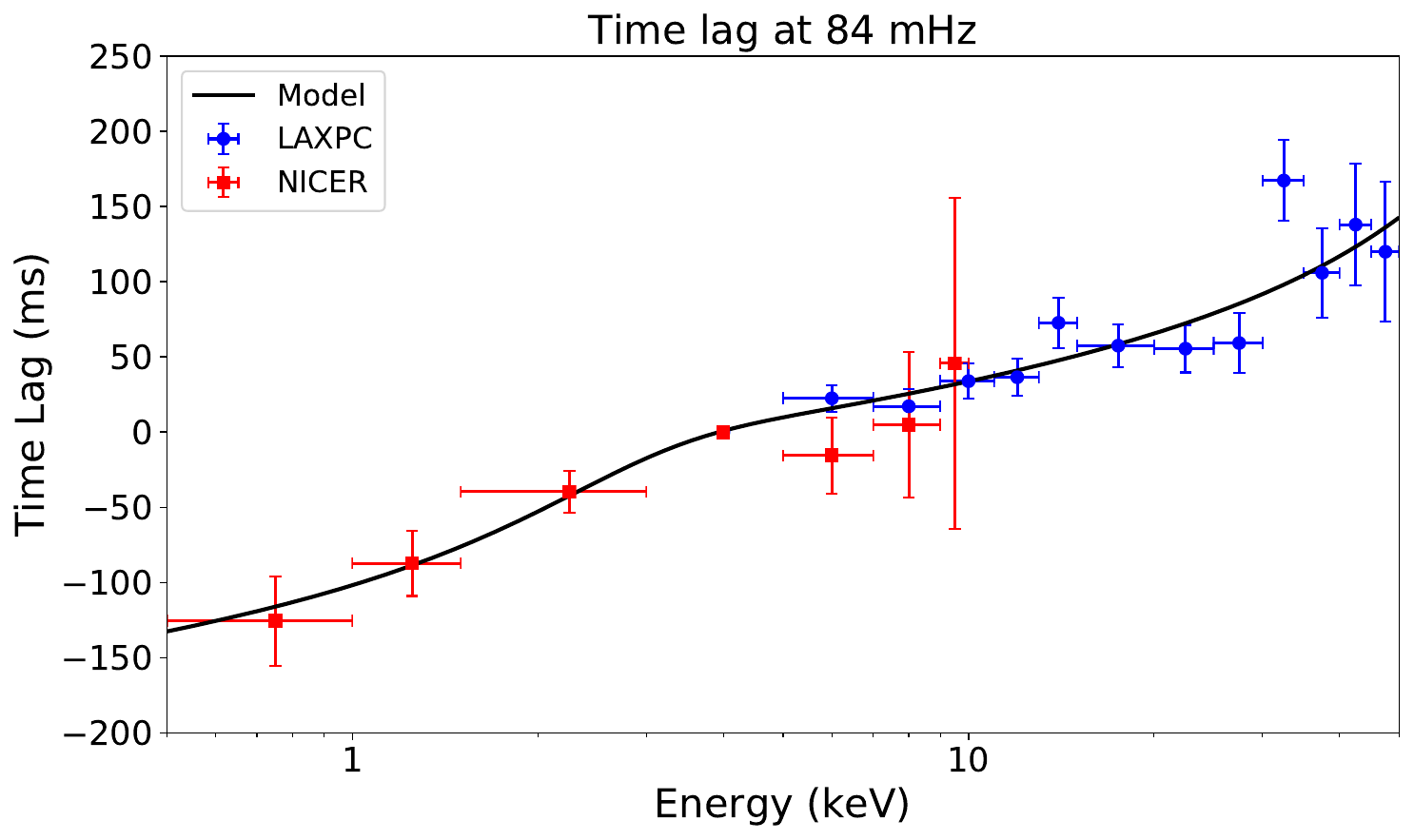}
\includegraphics[width=0.32\textwidth]{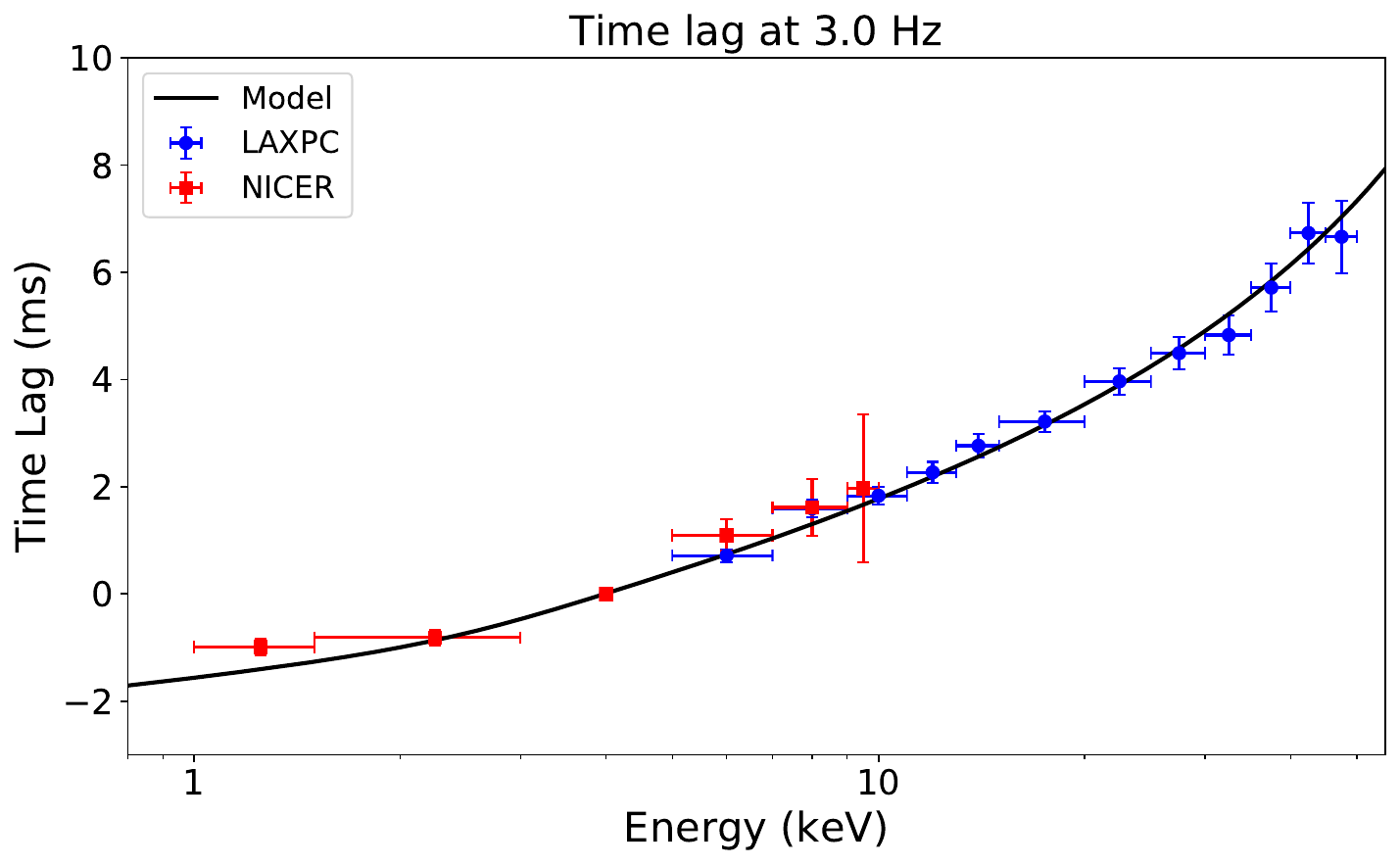}

\includegraphics[width=0.32\textwidth]{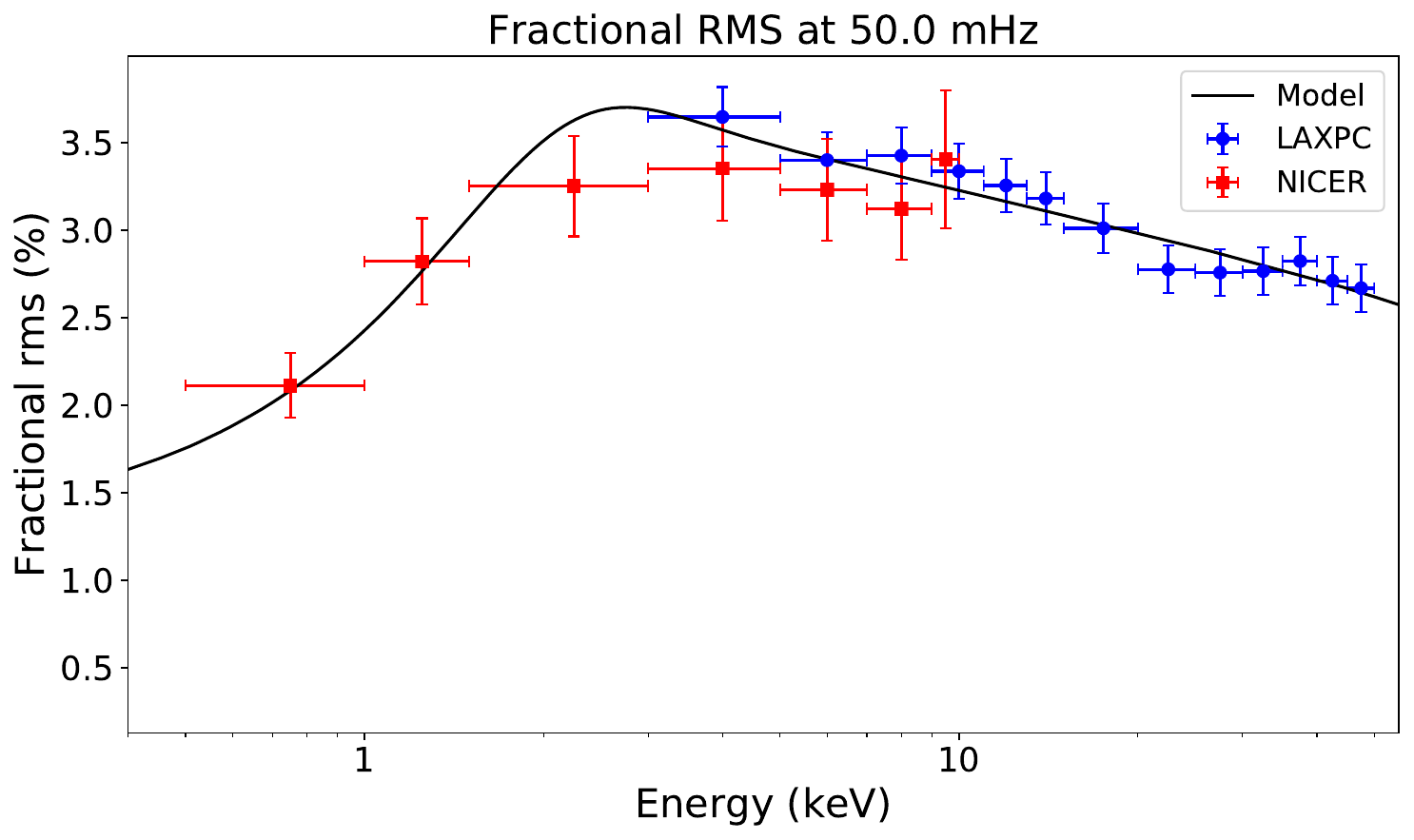}
\includegraphics[width=0.32\textwidth]{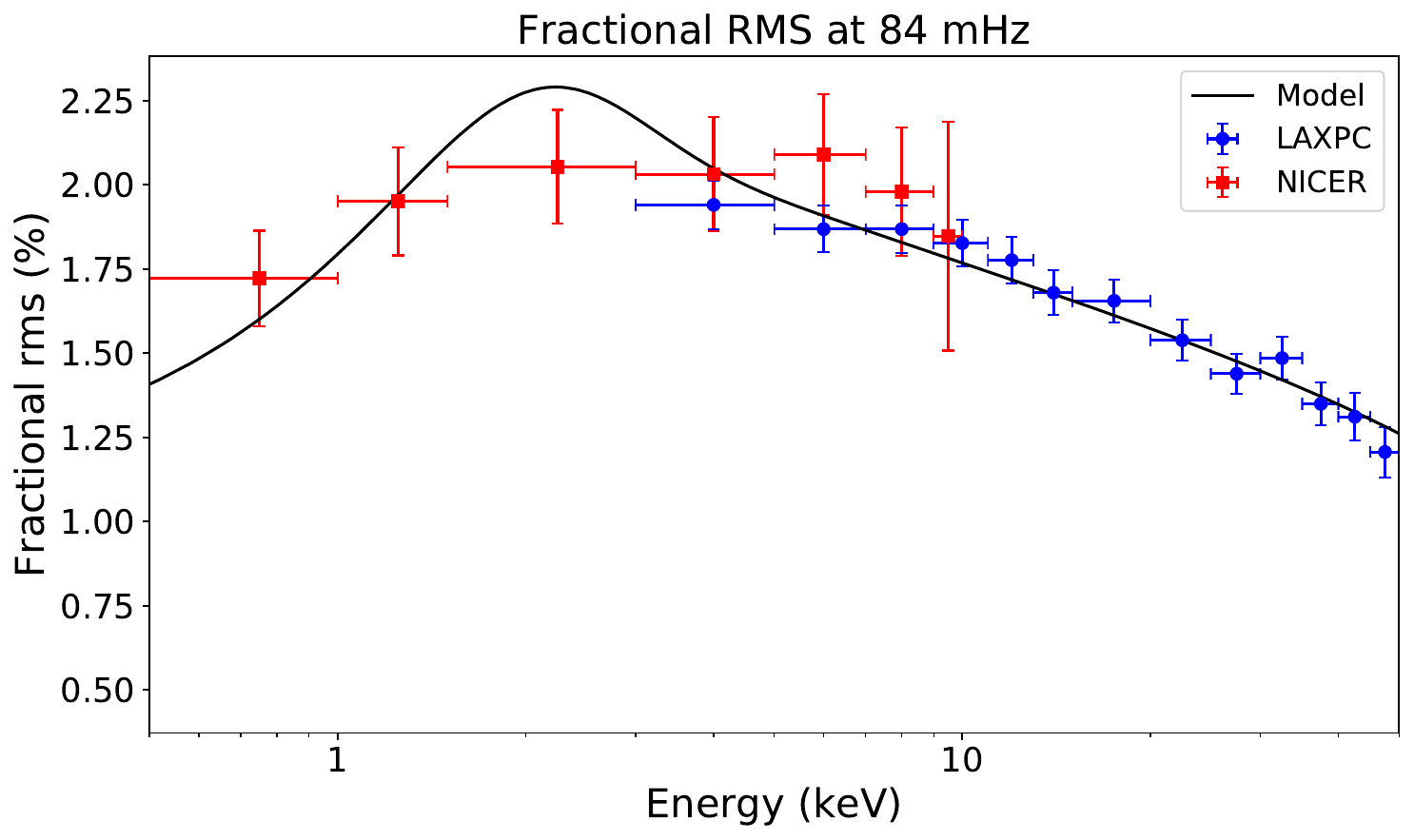}
\includegraphics[width=0.32\textwidth]{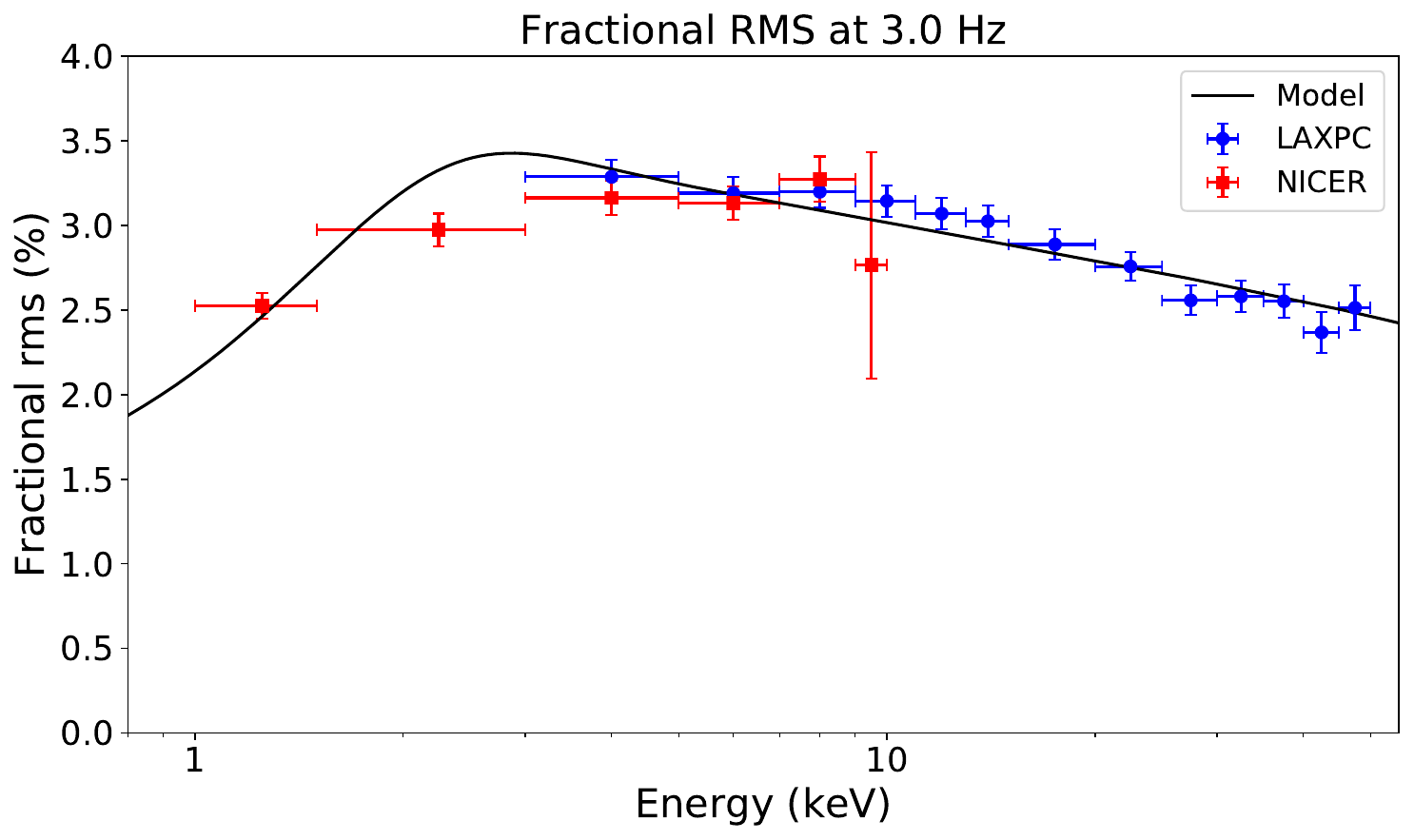}

\caption{Fractional rms (bottom) and time lag (top) as a function of photon energy at 50.0 mHz (left panel), 84 mHz (middle panel) and 3 Hz (right panel) frequencies from the nearly simultaneous NICER (N1 and N2; red filled circle) and AS1 (blue filled circle)  observations. The reference energy band used is the common range of 3-5.0 keV. The black solid curve represents the model fit derived from the model suggested by \cite{Garg_2020}.}
\label{fig:rms_lag}
\end{figure*}

In the first epoch, we extracted the frms and time lag at three characteristic frequencies (50~mHz, 84~mHz and 3~Hz) mentioned in Section  \ref{time}. As we can see, frms value increases as the energy increases in the broadband energies for the three frequencies, reaches a peak value, then decreases marginally in the higher energies (see Figure \ref{fig:rms_lag}). In the case of time lag, it increases with increasing energy. The frms and time lag spectra were fitted using \textsc{xspec}, following the same methodology as \cite{Garg_2020}. The modeling was started with $\phi_{\dot{H}}$ set to zero as a starting point, so we can compare and interpret the phases of other components relative to this. Initially, we modeled the observed variability by considering the scenario in which the variations arise solely from the changes in the accretion rate (parameterized through the inner disk temperature) and the coronal heating rate (derived from the electron temperature) and a phase difference between them $\phi_{kT_{\rm in}}$. This provided a poor fit, with reduced $\chi^2$ $>$ 2 at 50~mHz, 84~mHz, and 3~Hz. Thus, we added variations in the scattering fraction as a model parameter along with its corresponding phase. This model component improved the fit and provided a statistically acceptable fit. Further, we checked the model by adding other components, such as variations in the inner disk radius and optical depth, to variations in the accretion rate, the coronal heating rate, and the scattering fraction, but this did not improve the fit ($\chi^2 > 1.5$ at all three frequencies). Thus, we modeled the observed energy dependent variability properties by making use of the three different parameters namely, variation in inner disk temperature ($\delta kT_{\rm in}$), heating rate ($\delta\dot{H}$) and scattering fraction ($\delta f_{\rm sc}$) and the phase differences between them ($\phi_{kT_{\rm in}}$, $\phi_{\dot{H}}$, $\phi_{f_{\rm sc}}$). The best-fit values of the model parameters are given in Table \ref{table:rmslagfit}.

\begin{figure}
  \centering
  \includegraphics[width=1.0\linewidth]{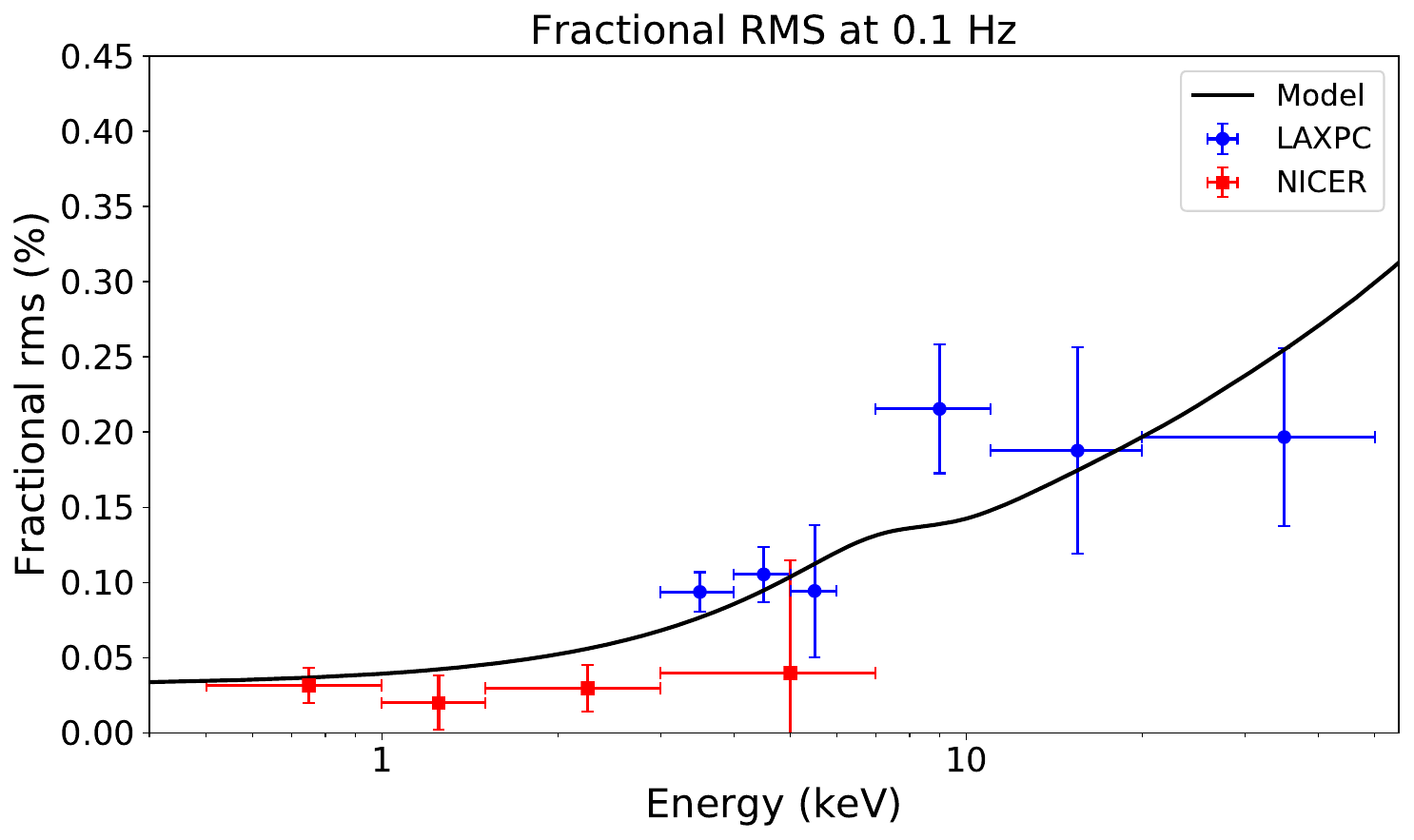}
   \caption{The frms calculated in the frequency 0.1~Hz for NICER and LAXPC data fitted with the model. The best fit values are summarized in Table \ref{table:rmslagfit}.}
\label{pt1_rms}
\end{figure}

\begin{table*}
\centering
\caption{Best fit values of parameters obtained by fitting the fractional rms and time lag spectra for the observed features at 50~mHz (QPO), 84~mHz and 3~Hz using the model described in \cite{Garg_2020}. All the error bars are within 90\% confidence level. Here,  $|\delta kT_{\rm in}|$, $|\delta \dot{H}|$, and $|\delta f_{\rm sc}|$ represent the variations in disk temperature, heating rate, and scattering fraction, respectively. $\phi_{\dot{H}}$ and $\phi_{f_{\rm sc}}$ represent the phases of $\delta \dot{H}$ and $\delta(f_{\rm sc})$ with respect to $|\delta kT_{\rm in}|$.}

\begin{tabular}{lcccc}
\hline
& \multicolumn{3}{c}{Epoch 1} & Epoch 2\\
\cline{2-5}
Parameters & QPO (50 mHz) & Harmonic (84 mHz) & Broad Feature (3 Hz) & 0.1 Hz\\
\hline
$|\delta kT_{\rm in}|$ (\%) & $0.81^{+0.09}_{-0.09}$ & $0.61^{+0.06}_{-0.06}$ & $0.75^{+0.06}_{-0.06}$ & $1.65^{+0.28}_{-0.28}\times10^{-2}$\\[6pt]
$|\delta \dot{H}|$ (\%) & $2.82^{+0.15}_{-0.15}$ & $1.90^{+0.13}_{-0.13}$ & $2.65^{+0.11}_{-0.11}$ & $3.96^{+1.30}_{-1.30}\times10^{-1}$\\[6pt]
$|\delta f_{\rm sc}|$ (\%) & $2.21^{+0.44}_{-0.45}$ & $0.68^{+0.29}_{-0.29}$ & $2.17^{+0.26}_{-0.27}$ & $-$\\[6pt]
$\phi_{kT_{\rm in}}$ & $-0.031^{+0.01}_{-0.01}$ & $-0.13^{+0.02}_{-0.02}$ & $-0.15^{+0.01}_{-0.01}$ & $0.0^{+0.82}_{-0.71}$\\[6pt]
$\phi_{\dot{H}}$ & $0.0^{f}$ & $0.0^{f}$ & $0.0^{f}$ & $0.0^f$\\[6pt]
$\phi_{f_{\rm sc}}$ & $0.0^{f}$ & $0.087^{+0.18}_{-0.08}$ & $=\phi_{kT_{\rm in}}$ & $-$\\[6pt]
$T_{kT_{\rm in}}$ (msec) & $98.7$ & $248.33$ & $8.2$ & $-$\\[6pt]
$T_{f_{\rm sc}}$ (msec) & $-$ & $164.92$ & $8.2$ & $-$\\
\hline
$\chi^2$/dof & $18.87/34$ & $23.39/33$ & $34.31/31$ & $11.73/9$\\
\hline
\end{tabular}
\label{table:rmslagfit}
\end{table*}

For the second epoch, the energy-dependent variability properties were extracted at 0.1~Hz, where the power in the PDS is relatively high (see Figure~\ref{PDS_AS2}). Unlike the hard state, the fractional rms increases monotonically with energy across the broadband range, without the peak and marginal decline seen at higher energies in the first epoch. The time lag is consistent with zero, with large uncertainties in the epoch, hence we did not model the time lag spectrum. The modeling started with variations in the heating rate alone, which yielded a poor fit ($\chi^2 \sim 104.66/10$). Adding the inner disk temperature as a second parameter significantly improved the fit ($\chi^2 \sim 11.67/9$), and no further parameter additions were pursued given the limited degrees of freedom. The best-fit values are given in Table~\ref{table:rmslagfit}.

\section{Discussion}
\label{conclu}

We have performed the spectro-timing analysis of MAXI~J1820+070 using nearly simultaneous NICER and AstroSat observations. The broadband (XTI and LAXPC in the 0.7 - 30 keV band) energy spectra from the different epochs fitted with a model combination that involves a disk emission and a Comptonizing component: \textit{constant * tbabs * edge * edge (thcomp * diskbb + gaussian)}. In the first epoch, the yielded spectral parameters ($kT_{in} \sim 0.38$~keV, $f_{sc} \sim 0.54$) and the spectral shape suggest that the source was in the hard spectral state of black hole X-ray binaries \citep{Remillard_2006, Belloni_2009}. In the second epoch, the observed spectra have a distinct shape compared to those in the first epoch, and the spectral parameters, such as $kT_{in} \sim 0.75$~keV and $f_{sc} \sim 0.03$, changed significantly. These results indicate that the source was in the high soft state. The spectral state information is consistent with previous studies of MAXI~J1820+070 \citep{Geethu_Prabhakar, Banerjee_2024}. In both hard and soft states, we observed the presence of a 6.4~keV Iron K$\alpha$ line feature in the broadband spectra.

In the hard state observation, the source exhibited three features in the PDS at 50~mHz, 84~mHz and 3~Hz. The sharp feature at 50 mHz is reported as a QPO in the previous study \citep{Mudambi_2020}. In that work, the authors used only AstroSat LAXPC and extracted frms and time lag spectra at representative frequencies such as 0.1, 1, and 10 Hz to understand the energy-dependent variability properties. In addition, they modeled these properties using a single stochastic zone propagation model \citep{Barimaqbool_2019} limited to the LAXPC band. In this work, we extracted the energy-dependent properties at 50~mHz, 84~mHz, and 3~Hz and modeled them using a sophisticated model described in \cite{Garg_2020}, which incorporates the different emission components. Modeling the energy-dependent variability properties provides a better understanding of the underlying mechanism for different variability components in the PDS. Across the three frequencies, we observed a common trend: the frms increases with energy, reaches a peak, then decreases marginally at high energies. The time lag increases with energy, suggesting a hard time lag at all three frequencies. 
\subsection{Feature at 50mHz}
At 50 mHz, we observed a negative phase of the inner disk temperature ($\phi_{kT_{in}} \sim -0.031$) with respect to the other components ($\phi_{\dot{H}}$ and $\phi_{f_{sc}}$), suggesting that the variations originate in the inner disk region. In addition, the perturbations took $\sim 99$ ms to reach the Comptonizing region from the disk.  Thus, the modeling suggests that variations that originate in the inner disk lead to simultaneous changes in the scattering fraction and the heating rate in the corona. 
\subsection{Feature at 84mHz}
Similarly, for the 84 mHz feature, perturbations started from the inner disk temperature ($\phi_{kT_{in}}=-0.131$), caused a change in the heating rate ($\phi_{\dot{H}}=0.0 $), then causing changes in scattering fraction ($\phi_{f_{sc}}=0.087$) after a small time delay. The modeling suggests that the perturbations from the inner disk took $\sim$ 165~ms to change in the scattering fraction, but the variation in the heating rate happened after $\sim 248$~ms compared to inner disk temperature variation (See Table \ref{table:rmslagfit}).
\subsection{Feature at 3Hz}
The broad feature around 3~Hz showed an frms and lag that could be modeled using a perturbation that originates simultaneously in the inner disk temperature and the scattering fraction ($\phi_{kT_{\rm in}} = \phi_{f_{\rm sc}} = -0.15^{+0.01}_{-0.01}$), with both varying in phase with no offset between them. These perturbations subsequently propagate towards the heating rate ($\phi_{\dot{H}} = 0.0$), which responds last. The perturbation travel timescale associated with both the inner disk temperature and the scattering fraction is found to be $T_{kT_{\rm in}} = T_{f_{\rm sc}} \sim 8.2$~ms, which is significantly shorter than 
the timescales found at 47~mHz and 84~mHz, consistent with the higher frequency of this feature. The percentage variations in the heating rate ($|\delta \dot{H}| = 2.65^{+0.11}_{-0.11}$\%) and the scattering fraction ($|\delta f_{\rm sc}| = 2.17^{+0.26}_{-0.27}$\%) are considerably larger than that of the inner disk temperature ($|\delta kT_{\rm in}| = 0.75^{+0.06}_{-0.06}$\%), suggesting that while the perturbation originates in the disk, the coronal parameters dominate the observed variability at this frequency, consistent with the trend seen at 47~mHz and 84~mHz. A general pattern is observed in the modeling of frms and lag spectra, that the variations originate from the inner disk. However, the calculated perturbation travel time for three features (QPO, the 84 mHz feature, and the broad 3.0 Hz feature) varies drastically. The modeling suggests that variations in both the soft and hard components are essential for reproducing the observed energy-dependent variability. The significant improvement in the fit upon inclusion of the scattering fraction indicates that the Comptonizing region plays a more dominant role in driving the variability than the inner disk radius \citep{Kara_2019}. In addition, the modeling provided insights into perturbation travel time and its dependence on frequency.
\\
\subsection{Feature at 0.1Hz in Soft State}
At 0.1~Hz in the soft state, the fractional rms is described by variations in the inner disk temperature and the heating rate, with the heating rate showing a considerably larger variation than the disk, consistent with the coronal dominance seen across all three features in the hard state. The frms amplitude in this epoch is overall significantly lower compared to the hard state, consistent with the known suppression of broadband variability in the soft spectral state \citep{Titarchuk_2008}. We note that the phase of the inner disk temperature ($\phi_{kT_{\rm in}} = 0.0^{+0.82}_{-0.71}$) is poorly constrained, and the result should be treated with caution, since only the frms and not the time lag could be modeled for this observation.
\subsection{Comparison with previous studies and other BHXRBs}
\citet{Mudambi_2020} investigated the energy-dependent variability of MAXI~J1820+070 using AstroSat/LAXPC data alone, employing the single-zone stochastic propagation model of \citet{Barimaqbool_2019}. In that formulation, the variability is described in terms of the normalized amplitude of the truncated disk temperature ($\delta T_s$), the hot inner flow temperature ($\delta T_e$), and the time lag between them ($\tau_D$). The time lag between the disk and the comptonizing region was found to be $\sim$100~ms at frequencies below 2~Hz. Our result at 47~mHz, where the perturbation travel timescale is $\sim$98.7~ms, is broadly consistent with this finding, confirming the frequency-dependent nature of the propagation timescales. However, the present analysis reveals an additional dependence on the scattering fraction, $f_{\rm sc}$, which is not a free parameter in the \citet{Barimaqbool_2019} model. The percentage variation in the scattering fraction ($|\delta f_{\rm sc}| = 2.21$\%) is comparable to that in the heating rate ($|\delta \dot{H}| = 2.82$\%), indicating that the comptonizing region contributes significantly to the observed variability at this frequency. The LAXPC data alone can be adequately described by variations in a limited set of parameters, such as the disk temperature and coronal heating rate, similar to the approach adopted by \citet{Mudambi_2020}. The inclusion of soft band data from NICER, in addition to LAXPC coverage, proved essential for constraining the frequency dependence, as the scattering fraction predominantly affects the softer part of the spectrum.

Several studies have employed the \citet{Garg_2020} formulation to investigate energy-dependent variability in black hole X-ray binaries, and a comparison of those results with the present work offers useful physical insight. The QPO at $\sim 47$~mHz is the most prominent feature in the PDS of MAXI~J1820+070 during the hard state observation. The modeling of the energy-dependent fractional rms and time lag spectra at this frequency, using the formulation of \citet{Garg_2020}, yields percentage variations (see Table \ref{table:rmslagfit}) of $|\delta kT_{\rm in}| = 0.81^{+0.09}_{-0.09}$\%, $|\delta \dot{H}| = 2.82^{+0.15}_{-0.15}$\%, and $|\delta f_{\rm sc}| = 2.21^{+0.44}_{-0.45}$\%. The phase of the inner disk temperature with respect to the heating rate is found to be $\phi_{kT_{\rm in}} = -0.031^{+0.01}_{-0.01}$, while the phase of the scattering fraction is fixed at $\phi_{f_{\rm sc}} = 0.0$. The negative value of $\phi_{kT_{\rm in}}$ indicates that the inner disk temperature leads the other components, implying that the perturbations originate in the inner disk and subsequently drive changes in both the scattering fraction and the coronal heating rate. The perturbation travel timescale from the disk to the
comptonizing region, $T_{kT_{\rm in}}$, is found to be $\sim$98.7~ms. \citet{Nazma_2023} performed an energy-dependent variability analysis of the QPO at 0.60~Hz in H~1743$-$322 using AstroSat data with \citet{Garg_2020} framework. In that source, the variability associated with the QPO was found to originate in the coronal heating rate, which subsequently drives changes in the inner disk temperature, a sequence that is opposite to what we find for MAXI~J1820+070, where the perturbations begin in the inner disk. This difference in propagation direction suggests that the dominant driver of QPO variability may not be universal across black hole X-ray binaries and may depend on the accretion geometry or the spectral state of the source at the time of observation. It is worth noting, however, that both sources show a hard time lag at the QPO frequency, indicating that the general character of the energy-dependent lag is similar even when the propagation sequence differs. Although the perturbations are found to originate in the inner disk, the percentage variation in the inner disk temperature ($|\delta kT_{\rm in}| = 0.81$\%) is considerably smaller than those of the coronal heating rate ($|\delta \dot{H}| = 2.82$\%) and the scattering fraction ($|\delta f_{\rm sc}| = 2.21$\%). This suggests that while the inner disk acts as the site of origin of the perturbation, the coronal parameters respond more strongly, indicating that the corona plays a dominant role in shaping the observed variability at the QPO frequency. A similar dominance of coronal variability over disk variability has been noted in other black hole X-ray binaries \citep{Kara_2019}, further supporting the picture in which the comptonizing region is the primary contributor to the observed variability at this frequency.

The harmonic feature at $\sim 84$~mHz is modeled with variations of $|\delta kT_{\rm in}| = 0.61^{+0.06}_{-0.06}$\%, $|\delta \dot{H}| = 1.90^{+0.13}_{-0.13}$\%, and $|\delta f_{\rm sc}| = 0.68^{+0.29}_{-0.29}$\%. The phase of the inner disk temperature is $\phi_{kT_{\rm in}} = -0.131^{+0.02}_{-0.02}$, and the phase of the scattering fraction is $\phi_{f_{\rm sc}} = 0.087^{+0.18}_{-0.08}$. The perturbation travel timescales are $T_{kT_{\rm in}} \sim 248.3$~ms and $T_{f_{\rm sc}} \sim 164.9$~ms. The negative phase of the inner disk temperature indicates that, as with the QPO, the perturbations at 84~mHz also originate in the inner disk. However, the propagation sequence differs from the QPO: at 84~mHz, the perturbations from the inner disk first drive a change in the scattering fraction after $\sim$164.9~ms, and subsequently produce a variation in the heating rate after $\sim$248.3~ms. This contrasts with the QPO at 47~mHz, where changes in the scattering fraction and the heating rate occur simultaneously. The longer travel timescale at 84~mHz compared to 47~mHz, despite the higher frequency of the harmonic, points to a more complex propagation path, possibly involving an intermediate interaction between the disk and the corona before the heating rate is affected. The AstroSat observations of H~1743$-$322 analysed by 
\citet{Nazma_2023} correspond to the hard spectral state, broadly comparable to the hard state of MAXI~J1820+070 studied here. In addition, both sources were observed at the peak of the outburst. In both sources, type-C QPOs were found at frequencies of 0.60~Hz (H~1743$-$322) and 47 mHz (MAXI~J1820+070). For the harmonic, \citet{Nazma_2023} found that the propagation sequence is reversed relative to the QPO, with the variability originating in the inner disk and subsequently driving changes in the heating rate. This is qualitatively consistent with our result for the 84~mHz harmonic of MAXI~J1820+070, where the perturbations likewise originate in the inner disk. The key difference lies in the involvement of the scattering fraction: in the present work, the scattering fraction plays an intermediate role in the propagation chain between the inner disk and the heating rate, whereas this parameter was not explicitly included in the analysis of \citet{Nazma_2023}. The observed time lags in H~1743$-$322 were found to be too large to be explained by light-crossing time alone, and a precession model incorporating reflection and reprocessing was suggested as a viable alternative. In our case, the perturbation travel timescales of $\sim$165-248~ms at 84~mHz are similarly large, suggesting that light-crossing time is unlikely to account for the delays and that propagation of accretion rate fluctuations through the disk-corona system is a more plausible explanation.

The broad noise feature near 3~Hz is modeled with variations of $|\delta kT_{\rm in}| = 0.75^{+0.06}_{-0.06}$\%, $|\delta \dot{H}| = 2.65^{+0.11}_{-0.11}$\%, and $|\delta f_{\rm sc}| = 2.17^{+0.26}_{-0.27}$\%. The phase of the inner disk temperature is $\phi_{kT_{\rm in}} = -0.154^{+0.01}_{-0.01}$, while the scattering fraction is found to originate simultaneously with the inner disk temperature, with $\phi_{f_{\rm sc}} = \phi_{kT_{\rm in}}$, indicating no phase offset between the two. Both the inner disk temperature and the scattering fraction vary in phase and together drive changes in the heating rate. The perturbation travel timescales for both $kT_{\rm in}$ and $f_{\rm sc}$ are found to be $T_{kT_{\rm in}} = T_{f_{\rm sc}} \sim 8.2$~ms, which is significantly shorter than the timescales found at 47 mHz and 84 mHz, consistent with the higher frequency of this feature. \citet{Tanenia_2024} observed GX~339$-$4 during the low/hard state of its 2021 outburst, which is comparable to the hard state epoch of MAXI~J1820+070. The 2.94~Hz broad feature in GX~339$-$4 is the closest in frequency to the 3~Hz feature in MAXI~J1820+070 among all the sources compared in this work, making this the most physically analogous comparison in the present discussion. Using the \citet{Garg_2020} formulation with variations in four spectral parameters, they modelled the energy-dependent fractional rms and time-lag spectra associated with broad features at 0.22~Hz and 2.94~Hz in the PDS. In that work, modeling with variations in disk normalization ($N_{\rm dbb}$) and heating rate alone proved insufficient, as observed features were found to carry an additional dependence on the scattering fraction. The variability was interpreted as originating in the inner disk temperature, subsequently propagating inward and driving changes in the disk normalization, scattering fraction, and heating rate in sequence. In our analysis of the broad $\sim 3$ Hz feature in MAXI~J1820+070, we similarly find that the scattering fraction is required to reproduce the observed variability, and that the inner disk temperature plays a central role. The key difference is that, in our case, the inner disk temperature and the scattering fraction originate simultaneously with no phase offset between them, while in GX 339$-$4 the scattering fraction responds after the disk, implying a different coupling between the disk and the corona in the two sources. However, in both cases, the heating rate is the last parameter to respond, suggesting that coronal heating is a consequence of, rather than a driver of, the observed variability at broad noise frequencies. \citet{Dhaka_2024} analysed GRS~1915+105 using 2017 AstroSat and NICER observations, where the source is identified in two different flux states. The broad feature at $\sim$2~Hz (present in both low and high flux states) is broadly comparable with the 3~Hz feature. In GRS~1915+105, variations in coronal parameters alone proved insufficient, and a satisfactory fit requires simultaneous variations in the inner disk temperature, the inner disk radius, and the coronal heating rate. The inner disk radius and temperature were found to be anti-correlated with no phase lag between them, while the negative phase lag of the coronal heating rate indicated that coronal heating variation precedes the disk response, a sequence that is notably different from what we find for MAXI~J1820+070, where the inner disk leads the coronal response. In our analysis, disk normalization ($N_{\rm dbb}$) was not required to model observed variability; the combination of inner disk temperature, scattering fraction, and heating rate was found to be sufficient. This suggests that the variability in MAXI~J1820+070 at this frequency is not driven by changes in the inner disk radius, consistent with the interpretation of \citet{Kara_2019}, who argued that coronal variability rather than disk truncation radius variations is the dominant driver in this source.
\\

Across all three variability features studied in this work, the perturbations are found to originate in the inner disk and propagate inward. The perturbation travel timescales decrease systematically from $\sim$248~ms (at 84~mHz) to $\sim$8~ms (at 3~Hz). This result is consistent with the expectation that higher-frequency variability is associated with perturbations arising at smaller disk radii. In all three features, although the disk is the origin of the perturbations, the fractional variations in the coronal parameters (the heating rate and the scattering fraction) are much larger, indicating that the corona plays a dominant role in shaping the observed energy-dependent variability. The broad energy coverage of the combined NICER and AstroSat data demonstrated that the scattering fraction is an essential model component for explaining the energy-dependent variability across all three frequencies. The propagation sequence, however, is not identical across frequencies. At 47~mHz, the heating rate and scattering fraction respond simultaneously to the disk perturbation, whereas at 84~mHz the scattering fraction responds first and the heating rate subsequently. However, at 3~Hz, the inner disk temperature and scattering fraction vary in phase with no offset between them. This frequency-dependent variation in the propagation sequence suggests that the coupling between the disk and the corona in MAXI~J1820+070 is not a simple, uniform process but evolves with the characteristic timescale of the variability feature.
\\
Our results broadly consistent with two previous lag studies of the source. \cite{kawamura_2023}  identified a QPO  and harmonic from Insight-HXMT at 36.6~mHz and 94.4~mHz, respectively, which are close to the $\sim$47~mHz and $\sim$84~mHz features identified in AstroSat/NICER data. They modelled broadband time lag with a propagating-fluctuations framework. The lag between the soft and hard Comptonization regions is on the order of 50\,ms at lower frequencies. However, the lags decrease at higher frequencies as fluctuations are generated closer to the compact object. This trend is qualitatively consistent with our finding that the lag amplitude changes from $\sim$250~ms (at the lower frequency) to $\sim$8~ms (at the higher frequency). \cite{kawamura_2023} constrain the radial structure and propagation speed of the hot flow using a physical, radially stratified model fitted to the broadband PSDs and the phase lag between two energy bands. Our approach is complementary. The \cite{Garg_2020}  framework computes the energy-dependent rms and lag from variations in the spectral parameters. Here we apply it with \texttt{diskbb} and \texttt{thcomp} to map the variability onto physical parameters: the inner disk temperature, the scattering fraction and the coronal heating rate. Using the energy-resolved fractional rms and lag, we model the 47~mHz QPO, the 84~mHz harmonic and the $\sim$3~Hz broad feature individually, and determine which parameters drive each one, by how much, and in what order. Since our delay timescales are between model parameters rather than a propagation speed through the flow, the two sets of results address different aspects of the same variability and can be read together. \cite{Omama_2023}, using a time-domain differential cross-correlation technique on NICER data, reported a much larger lag ($\sim$3~s) between the disk and Comptonized emission. This difference reflects methodology rather than physical inconsistency, since their CCF-based measurement is dominated by the slowest, highest-amplitude fluctuations in the light curve, whereas our frequency-resolved analysis isolates the lag associated with each individual Lorentzian component. Both studies show the disk leading the corona, as expected for inward-propagating fluctuations.

\section*{Acknowledgements}

 We thank the anonymous referee for constructive comments that have improved the manuscript. We are grateful to the LAXPC and SXT teams at TIFR, Mumbai, for sharing the data through the ISSDC archive and providing the essential software tools. This research has made use of data, software and/or web tools obtained from the High Energy Astrophysics Science Archive Research Center (HEASARC), a service of the Astrophysics Science Division at NASA/GSFC and of the Smithsonian Astrophysical Observatory’s High Energy Astrophysics Division. We acknowledge the use of Grammarly to proofread the paper. MB thanks the Inter-University Centre for Astronomy and Astrophysics (IUCAA) visiting program. VJ acknowledges the support provided by the Department of Science and Technology (DST) under the ‘Fund for Improvement of S \& T Infrastructure (FIST)’ program (SR/FST/PS-I/2022/208). VJ also thanks the Inter-University Centre for Astronomy and Astrophysics (IUCAA), Pune, India, for the Visiting Associateship.

\section*{Data Availability}

The data utilized in this work are publicly available on the Indian Space Science Data Center (ISSDC) website
\url{http://astrobrowse.issdc.gov.in/astro_archive/archive}
for the AstroSat mission and the HEASARC database
\url{https://heasarc.gsfc.nasa.gov} for the NICER mission.

\bibliographystyle{aasjournalv7}
\bibliography{refer}



\end{document}